\documentclass[preprint]{elsarticle}
\usepackage[T1]{fontenc}
\usepackage[utf8]{inputenc}
\usepackage{hyperref}
\usepackage{amsmath}
\usepackage[euler]{textgreek}
\usepackage{nccmath}
\usepackage{float}
\usepackage{subcaption}
\usepackage{graphicx}
\usepackage{url}

\journal{Borsa Istanbul Review}

\biboptions{authoryear}

\begin{document}

\begin{frontmatter}

\title{An Empirical Study on Arrival Rates of Limit Orders and Order Cancellation Rates in Borsa Istanbul}

\author[first_address]{Can Yilmaz Altinigne}
\ead{can.altinigne@epfl.ch}

\author[second_address]{Harun Ozkan}
\ead{harun.ozkan@matriksdata.com}

\author[second_address]{Veli Can Kupeli}
\ead{velican.kupeli@matriksdata.com}

\author[third_address]{Zehra Cataltepe}
\ead{cataltepe@itu.edu.tr}

\cortext[mycorrespondingauthor]{Corresponding author}

\address[first_address]{School of Computer and Communication Sciences, EPFL, Switzerland}
\address[second_address]{Matriks Bilgi Dağıtım Hizmetleri, 34396, Istanbul, Turkey}
\address[third_address]{Department of Computer Engineering, Faculty of Computer and Informatics
Engineering, Istanbul Technical University, Istanbul, Turkey}

\begin{abstract}
    Order book dynamics play an important role in both execution time and price formation of orders in an exchange market. 
    In this study, we aim to model the limit order arrival rates in the vicinity of the best bid and the best ask price levels. 
    We use limit order book data for Garanti Bank, which is one of the most traded stocks in Borsa Istanbul. 
    In order to model the daily, weekly, and monthly arrival of limit order quantities, three different discrete probability distributions are tested: Geometric, Beta-Binomial and Discrete Weibull. 
    Additionally, two theoretical models, namely, Exponential and Power law are also tested. We aim to model the arrival rates in the first fifteen bid and ask price levels.
    We use $L_1$ norms in order to calculate the goodness-of-fit statistics.
    Furthermore, we examine the structure of weekly and monthly mean cancellation rates in the first ten bid and ask price levels.
\end{abstract}

\begin{keyword}
Order arrival processes \sep Probability distribution fitting \sep Limit order book \sep Queueing systems
\JEL C46\sep C51
\end{keyword}

\end{frontmatter}

\section{Introduction}

One of the main research area on high-frequency financial data is to investigate the microstructural properties of stock markets. Generally, the research on this area contains modeling the main characteristics of the limit order book (\cite{cont2010stochastic, bouchaud2002statistical, zovko2002power}) and the behavior of traders of stocks around specific events (\cite{mu2010order}).
%Using the 
%previous research on the limit order book, we aim to conduct an empirical research on arrival 
% rates of limit orders and the behavior of cancel orders. 

Modeling some market elements such as the duration between two orders, order volumes and order arrivals using parametric statistical 
distributions can help to understand the structure of market dynamics. Exponential family distributions 
are widely used in modeling these types of exchange market elements (\cite{cont2010stochastic, jiang2008scaling}). Alternatively, 
arrival rates of limit orders can be modeled using a power law (\cite{bouchaud2002statistical, zovko2002power}). By modeling the order dynamics in the market, we can have some basic insight about the market microstructure. For this purpose, we aim to model the arrival rates
of limit order in the exchange market. Also, we intend to observe the statistical features of the cancellation rates (ratios of cancel orders to outstanding orders).

We use three well-known discrete statistical distributions for modeling the arrivals of orders: Discrete Weibull, Geometric and Beta - Binomial. In addition, we 
fit Exponential distribution and Power law on the same variables. We use $L_1$ norms between probability 
mass functions of discrete distributions and the true arrival rates. For continuous distributions, 
we discretize the fit results by calculating the area below the probability density function. We compare the 
performance scores of different fits using Welch's \textit{t}-test.

After completing the discrete and continuous distribution fits on the arrival rates of limit orders, we compare
the best fitting discrete model which is Discrete Weibull model with the theoretical models. We show that the performance of Discrete Weibull model 
is three times better than the Exponential model in terms of $L_1$ norms, and it is very competitive against the Power law model which is suggested by \cite{bouchaud2002statistical}.
In this part of our research, we present that the arrival rates of limit orders in the vicinity of the best prices
(15 ticks or less) can also be represented by discrete models.

In addition to the analysis of limit orders, we conduct a research on cancel orders. We analyze the cancellation rates
on the weekly and monthly basis. We consider the first 10 bid and the ask price levels.
We investigate whether the behavior of order cancellation rates change with respect to different bid and ask price 
levels. In our research, we observe that the hypothesis which implies that weekly and monthly 
mean order cancellation rates are consistent with Uniform distribution can not be statistically
rejected. 

\section{Background and Literature Review}

In exchange markets, limit orders, market orders and cancel orders constitute the current market dynamics. Arrived limit orders in the market create a limit order 
book. Buy and sell orders are placed in the limit order book according to their price and the quantity. 
Until a market order or a cancel order is executed on a particular limit order, 
that limit order stays in the order book (\cite{cont2011statistical}). Cancel orders delete limit 
orders in the table. Market orders execute limit orders and carry out the buying and selling
operation in the market. The highest price on the buy side represents the bid price, and the lowest price on the sell side 
represents the ask price. The prices on the buy (sell) side of the limit order book are arranged in descending 
(ascending) order. The mean of the bid and the ask price is the mid-price. The ask price is always higher than 
the bid price only during continuous auction which is the phase that the continuous trading occurs in the market. 
Difference between them is named as the bid/ask spread (\cite{cont2010stochastic}). 

\begin{figure}[htp]

    \centering
    \includegraphics[width=.8\textwidth]{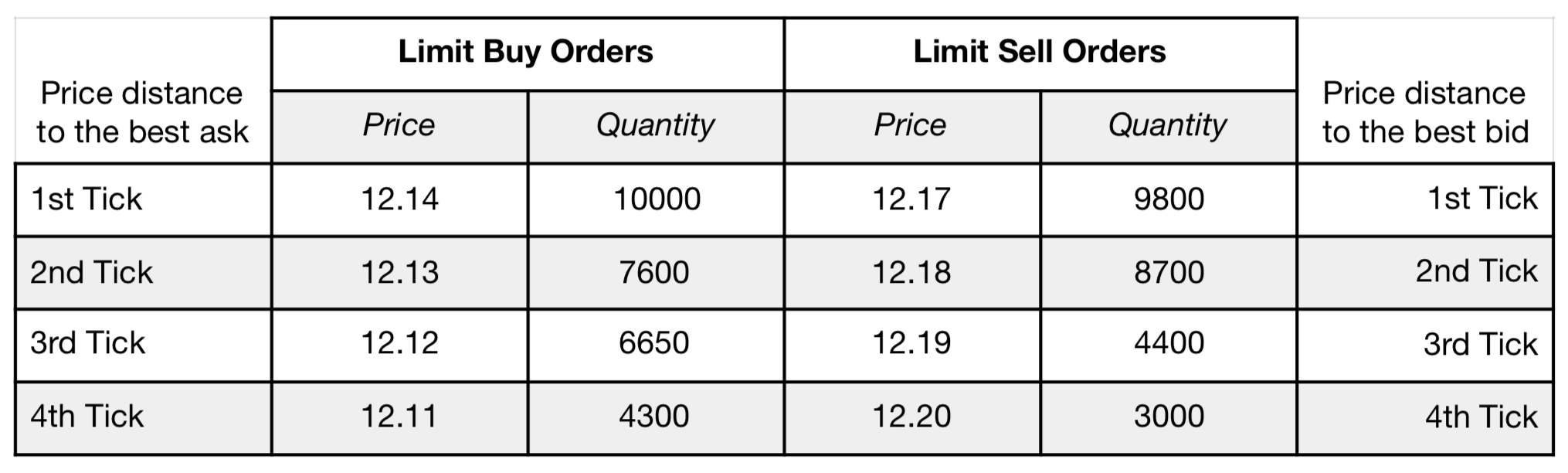}
    \caption{Example of a limit order book}
    \label{fig:}
    
\end{figure}

An example of a limit order book is shown in Figure 1. In this limit order book, the bid price is 12.14 and 
the ask price is 12.17. Limit buy (sell) orders represent the traders that would like to buy (sell) a quantity of a stock with 
the indicated price. If someone would like to buy (sell) a stock with a price equal or higher (lower) than 
the ask (bid) price, the operation is executed just after the submission of that particular limit buy (sell) 
order (\cite{cont2011statistical}).

The main part of our research contains the modeling of arrival rates of limit orders.
For limit buy (sell) orders, we consider the quantities on the left (right) side of the limit order book. The price levels
in the vicinity of the best prices are named as ticks. When we consider limit buy (sell) orders, we 
examine the order quantities in the ticks with respect to their distance to the best ask (bid) price (\cite{cont2010stochastic}).
For example, the price of 12.14 (12.17) is the first tick, and the price
of 12.13 (12.18) is the second tick for the limit buy (sell) orders. As shown in this 
example, the tick values are given to limit buy (sell) orders with respect to the order price's distance to the 
best ask (bid) price. Since a limit order book has a dynamic structure, the 
prices always change during the day as a result of incoming limit orders, execution of market and cancel 
orders. In our research, we model the order quantities in the first 15 ticks for both limit buy and 
limit sell order quantities.

The arrival of the limit orders near the best prices is dense. The price of an order is a crucial 
criterion of order execution, since the distance to the current bid and ask price is correlated with the  
rate of the arrival of limit orders (\cite{cont2010stochastic}). There are different remarks on explaining
arrival rates of limit orders mathematically. In previous research on order placement 
strategies, \cite{bouchaud2002statistical} and \cite{zovko2002power} suggested that arrival rates of limit orders $\Lambda (i)$ can be
modeled with a power law which can be seen below.

\begin{align}
    \Lambda (i) = \frac{k}{i^{\alpha}}
\end{align}

In this equation, the value \textit{i} 
denotes the tick value. The value of $\Lambda (i)$ can be between 0 and $\infty$. The $k$ parameter is a 
positive real number and $k$ and $\alpha$ can be estimated using a least squares fit 
as it is shown in Equation (2) (\cite{cont2010stochastic}). The value of 15 is chosen for an upper
boundary in that equation, since we perform modeling in the first 15 ticks. The real arrival rates 
are denoted with $\hat{\Lambda}(i)$.

\begin{align}
    \min_{k, \alpha} \sum_{n=1}^{15} (\hat{\Lambda}(i) - \frac{k}{i^{\alpha}})^2
\end{align}

In another research, a stochastic model consisting of independent Poisson processes is suggested 
by \cite{cont2010stochastic}. In this model, arrival rates of limit orders are modeled in such a way that they are distributed exponentially.

Weibull and \textit{q}-exponential distributions are used in an early work on modeling the duration between two successive transactions (\cite{jiang2008scaling}). In another work, the arrival of orders are represented 
as a renewal process where the waiting times between two successive orders are distributed according to 
Weibull distribution (\cite{cincotti2006waiting}). We can infer that the duration between two consecutive
orders would also be different in every tick, since the arrival rates of limit orders differ
with respect to the distance to the best prices. As a result, because of its flexibility in modeling
durations, we test Weibull distribution in modeling arrival rates of limit orders. However, we use the discrete variant of Weibull distribution 
proposed by \cite{nakagawa1975discrete}, since we perform an analysis on discrete distributions. Discrete Weibull distribution has 
two real number parameters $q > 0$ and $\beta > 0$ with integer support on [0, $\infty$). The probability mass function of Discrete Weibull can be seen in Equation (3)
(\cite{nakagawa1975discrete}). We use the probability density function as it starts from the least tick value of 1.

\begin{equation}
    P(X=x;q,\beta)=q^{(x-1)^{\beta}}-q^{x^{\beta}},\quad x=1,2,3,\ldots 
\end{equation}

Beta family distributions are frequently used in finance.
It is also often used in modeling the rates of recovery from debts and credit risk (\cite{chen2013curve}). It is well-known that high skewness is frequently observed in credit risks data (\cite{schroeck2002risk}). Similarly, arrival rates of limit orders have positive skewness since the rates are much higher in the ticks that are close to the best prices.
Because of its good performance in highly skewed data, we also test Beta Binomial distribution, which is a discrete member of Beta family distributions, in modeling the arrival rates.
Beta Binomial distribution is defined by two real number parameters: $\alpha > 0$ and $\beta > 0$. Both parameters have finite integer support on [0, $n$). The probability mass function for $n$ trails in Beta Binomial can be seen in Equation (4).

\begin{equation}
    P(X=x;\alpha,\beta) = \binom{n}{x}\frac{B(x + \alpha, n - x + \beta)}{B(\alpha, \beta)} 
\end{equation}
\begin{equation}
    \forall u, \forall v > 0, B(u,v) = \int_{0}^{1} t^{u-1} (1-t)^{v-1} dt
\end{equation}

As indicated before, Exponential distribution is also used for modeling arrival rates of limit orders.
In an early work on modeling the market dynamics, \cite{cont2010stochastic} assumed that limit orders arrive at the tick i from 
the best price with an exponential rate $\lambda(i)$ in his stochastic model. As a result, Exponential 
distribution and Geometric distribution which is the discrete variant of Exponential are also used in our research. Exponential 
distribution has one real number parameter $\lambda > 0$. The probability density 
function of Exponential distribution is given below.  

\begin{equation}
    P(X=x;\lambda) = \lambda e^{-\lambda x},\quad x \geq 0 
\end{equation}

Geometric distribution has one real number parameter $0 < p < 1$. The probability mass function of Geometric distribution for $x$ trails is given below.
We use the probability density function of Geometric distribution starting from the least tick value of 1. 

\begin{equation}
    P(X=x;p) = (1-p)^{x-1} p,\quad x=1,2,3,\ldots 
\end{equation}
Since most of the continuous and discrete distributions have support for the set $\{x: x>0 \}$, we adjusted our first tick rate to be the zeroth tick and, therefore, shifted the fit results to one tick right.

In an early work on cancel orders, \cite{blanchet2013continuous} assumed that the cancellation rates are relatively
higher in the ticks that are close to the best bid and ask prices than distant ticks.
\cite{cont2010stochastic} made an assumption that the cancellation rates show an Exponential distribution with respect to distance 
to the best bid and ask prices, and these rates are proportional to the limit orders in that level.
\cite{bouchaud2018trades} also suggested that the cancellation rates are proportional to the arrival rates of limit orders with an assumption 
that the activity is much higher in the area that has high arrival rates of limit orders.

\section{Materials and method}

\subsection{Market Data}
The pure market data contains network captured MoldUDP packets consisting of ITCH\textsuperscript{\textregistered} messeages. An ITCH\textsuperscript{\textregistered} NASDAQ protocol for market data
includes all orders in nano-second scale (\cite{nasdaq2015}). We use Garanti Bank stock data in Borsa Istanbul. The data spans 40 trading days from August 1, 2017 to September 29, 2017, and we sample 228 instances 
which contain the rates of daily, weekly and monthly arrived limit orders to analyze. 

\subsection{Arrival Rates of Limit Orders}
We extract the information of limit order quantities 
arrived at the first 15 ticks to the best prices. There are quantities arrived after the first 15 ticks, but 
they were few with respect to the quantities in the first 15 ticks so we omitted those quantities. As a result, 
we perform discrete and continuous fits on $\lambda^t (i)$ which indicates the density of quantities arrived $Q^t (i)$ at the $i$th 
tick in time instance $t$. 

\begin{equation}
    \lambda^t (i) = \frac{Q^t (i)}{\sum_{i=1}^{15} Q^t (i)},\quad i=1,\ldots,15 
\end{equation}

We split the data into four different time groups. These timesteps are the daily average of limit buy/sell quantities 
(40 days), the weekly average of limit buy/sell quantities (9 weeks), the monthly average of limit buy/sell quantities and
the hourly average of limit buy/sell quantities in 9 weeks. We consider the market working hours from 10 am to 1 pm
and 2 pm to 6 pm. We created 7 different hourly timesteps in a day. 

Consequently, we obtain 40 instances for daily data, 9 instances for weekly data, 2 instances 
for monthly data and 63 instances for hourly-weekly data. Since we consider both limit buy and limit sell orders,
we have 228 different instances to perform discrete and continuous fits. Using three discrete 
distributions, Discrete Weibull, Beta-Binomial and Geometric, we perform fits on the limit buy/sell order quantities that arrived at 
the first 15 ticks to the best prices. We used Exponential distribution as a continuous model 
approach (\cite{cont2010stochastic}). Also we compared the performance of the best discrete fit 
with Exponential fits and Power law fits which are proposed by \cite{bouchaud2002statistical} and \cite{zovko2002power} in order to examine if a discrete approach can compete with the approaches that 
are suggested in previous works.

Maximum likelihood estimation finds the parameters that maximize the joint probability density function 
of data (likelihood). Since the maximization is arduous for multiplication operation, 
in general, the logarithm of the likelihood function is considered (\cite{myung2003tutorial}). 
The approach of maximum likelihood estimation is shown in the Equation (10). In the equation $\theta$ is 
the parameter vector of the model, and $x_{1:n}$ is the data.

\begin{equation}
    Likelihood(\theta) = p(x_{1:n} \ | \ \theta) = \prod_{i=1}^{n} P(x_i \ | \ \theta) \ , \quad x_{1:n} = \left\{ x_1, x_2, \ldots , x_n \right\}
\end{equation}
\begin{equation}
    \hat\theta = \underset{\theta}{\operatorname{arg\,max}} \ p(x_{1:n} \ | \ \theta)
\end{equation}

As indicated before, we did not use any functions of \texttt{R} to estimate parameters of Exponential and 
Geometric distributions. When we take the derivative of the logarithm of the likelihood functions and 
equate it to zero, we can find the maximum likelihood estimation of the parameter $\lambda$ of Exponential 
distribution and the parameter $p$ of Geometric Distribution. 

\begin{equation}
    P(X=x;\lambda) = \lambda e^{-\lambda x}, \quad \geq 0 
\end{equation}
\begin{equation}
    \hat\lambda, \hat p = \frac{n}{\sum_{i=1}^{n} x_i}, \quad x_{1:n} = \left\{ x_1, x_2, \ldots , x_n \right\}
\end{equation}

The estimated parameter of Geometric distribution is also found using the same equation. Because Geometric 
distribution is the discrete variant of Exponential distribution, the only difference is that Geometric 
distribution has integer $x_{1:n}$ values. Estimation of the parameters of Exponential and Geometric distributions 
can be seen in Equation (12). 

In order to compare the performance of the models, we consider the sum of $L_1$ norms between the real values and the fit
results. Sum of absolute values of differences between observed densities and fit results are considered as 
the error term. The error term at timestep $t$ is shown below.

\begin{equation}
    \text{Error = } \sum_{i=1}^{15} \ \lvert \lambda^t (i) - \hat\lambda^t (i) \rvert
\end{equation}

\subsection{Order Cancellation Rates}

We analyze the number of arrived cancel orders around the best price and the ratio of cancel orders 
in the vicinity of the best prices. The ratios and numbers of cancel orders are considered on average 
weekly and monthly basis. We consider the quantity of the particular order and the total quantity in that tick
before that particular cancel order arrives. Then we sum these ratios on the monthly and weekly basis and divide the number of cancel 
orders that arrive in a particular tick on the monthly and weekly basis. We express the cancel order ratios in tick $i$ with $k$ 
arrived cancel orders in timestep $t$ with as $C_t(i)$.

\begin{align}
    C_t(i) = \frac{\sum_{n=1}^{k} \frac{\text{Canceled Quantity in tick $i$ with order $p_n$}}{\text{Total Quantity in tick $i$ before order $p_n$}}}{\text{Number of Cancel Orders Arrived in tick $i$ in timestep $t$}}
\end{align}

We compare our experiments on the number of cancel orders arrived in the vicinity of the best bid and ask prices and 
the behavior of the ratios of cancel orders with respect to the distance to the best bid and ask prices with 
previous works on cancel orders (\cite{cont2010stochastic, blanchet2013continuous, bouchaud2018trades}).

\section{Results}

\subsection{Discrete Fits on Arrival Rates of Limit Orders}

In order to find the performance of the discrete fits, we give each distribution fit a performance score.
The performance score is the ratio of the error of a distribution fit to the minimum fit error on that 
instance. As a result, this ratio is higher or equal to 1. If a distribution has the best fit, then its 
score becomes 1. We consider the performance according to the closeness of performance scores to 1. The equation 
of the normalized performance score of a distribution $d$ for instance $i$ is given below.

\begin{equation}
    NPS_d(i) = \frac{\text{Error of distribution $d$ on instance $i$}}{\text{Minimum error among 3 distributions on instance $i$}}
\end{equation}

We calculate the mean and standard deviation of performance scores of three distributions
in hourly, daily, weekly and monthly fits, and decide which distribution has the best fit in a particular timestep.
The limit buy and limit sell fits of Geometric, Beta-Binomial and Discrete Weibull distributions for the last 12 days (from Day 29 to Day 40) can be seen in Figure 2. 
There are fits for 40 days, but showing all of them might occupy a lot of space. Because of that, we 
only show fits of last 12 days.

\begin{figure}
    \centering
    \begin{subfigure}[b]{.92\textwidth}
        \includegraphics[width=\linewidth]{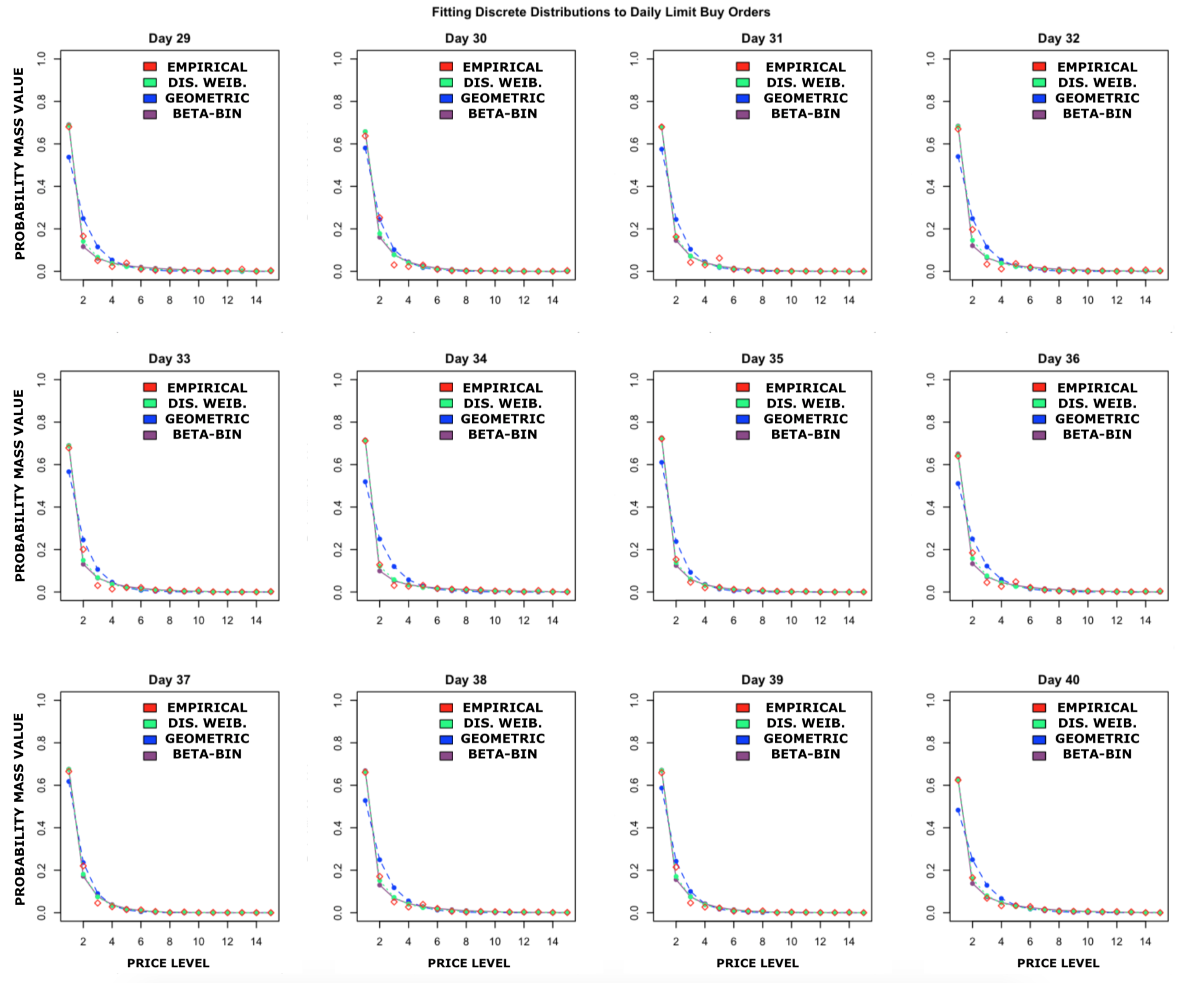}
    \end{subfigure}
    
    \begin{subfigure}[b]{.92\textwidth}
        \includegraphics[width=\linewidth]{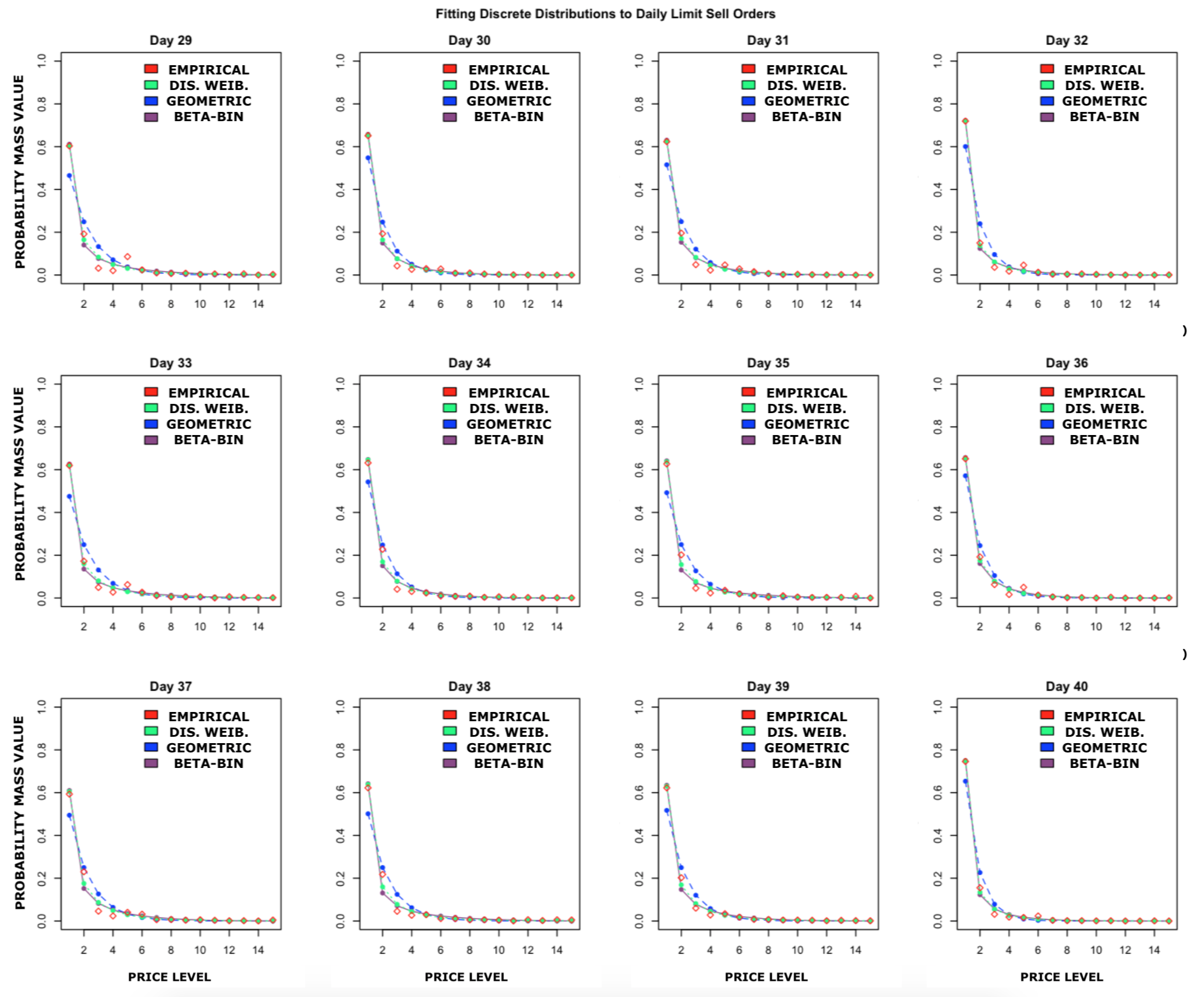}
    \end{subfigure}
    \caption[]{The upper figure shows Daily Limit Buy orders and the lower figure shows Daily Limit Sell orders}
\end{figure}

Day 29 and Day 30 are Thursday, September 14th and Friday, September 15th respectively.
Day 31 to Day 35 is the week starting on Monday, September 18th. Day 36 to Day 40 is the week starting on 
Monday, September 25th. Since some of the continuous and discrete probability distributions that we use have 
a support from 0 to infinity, we divide the probability mass values by the sum of all probabilities from the first 
tick to 15th tick for normalization. 
It is striking to observe that Discrete Weibull distribution has the best fits for 75 instances out of 80. Beta Binomial 
outperforms the fit performance of Discrete Weibull distribution for only 5 instances, and the fits of 
Geometric distribution is 3 to 4 times worse than both Discrete Weibull and Beta-Binomial distribution.

Normalized performance scores of each model in different time steps can be observed in the charts in Appendix section. The performance of the model increases as the cell color gets lighter and close to 1.

The limit buy and limit sell fits of Geometric, Beta-Binomial and Discrete Weibull distributions from Week 1 to Week 9 can be seen in Figure 3
and Figure 4 below. 

\begin{figure}[H]

    \centering
    \includegraphics[width=\linewidth]{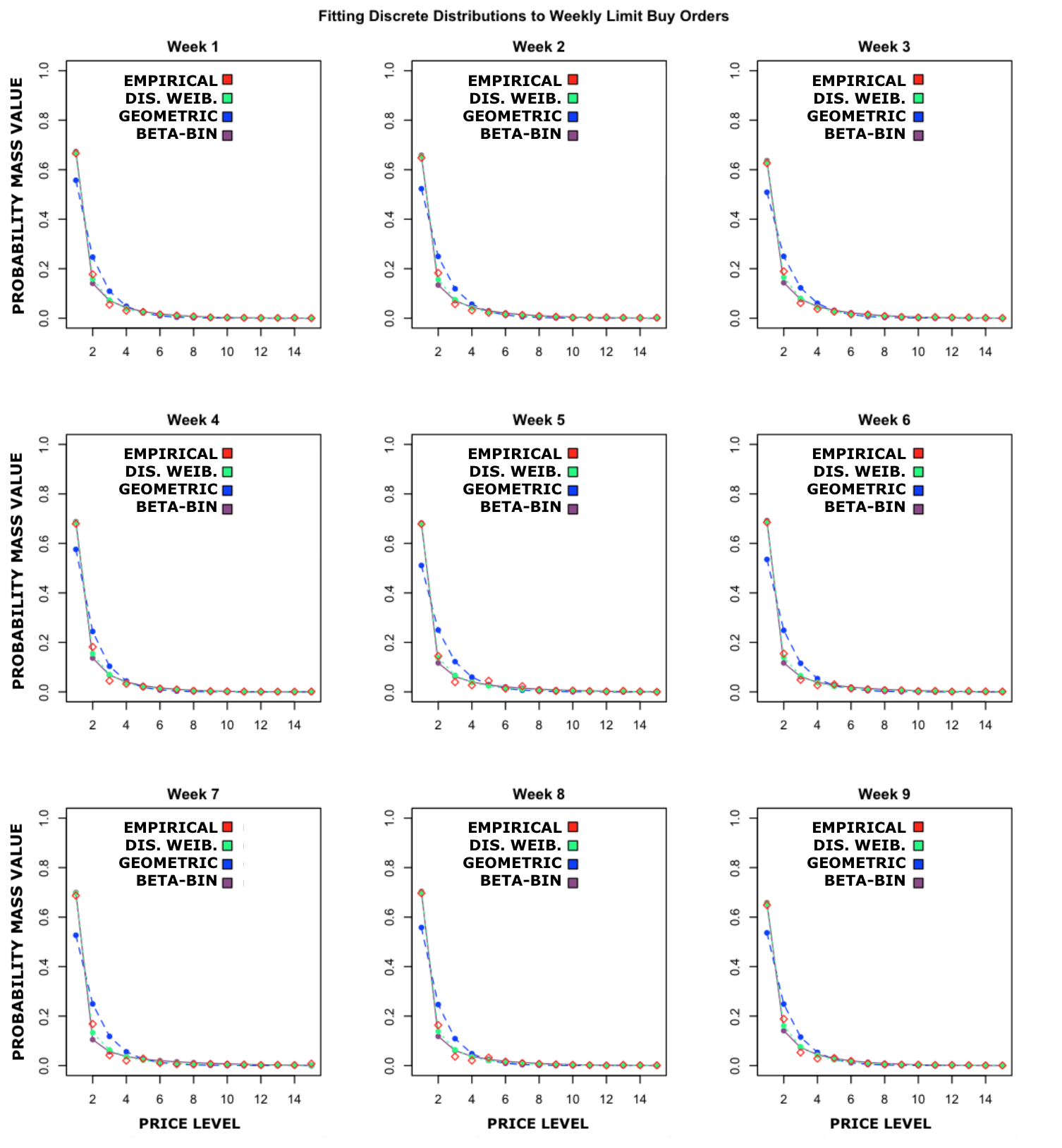}
    \caption{Discrete Fits on Weekly Limit Buy Arrival Rates}
    \label{fig:}
    
\end{figure}

\begin{figure}[!ht]

    \centering
    \includegraphics[width=0.7\linewidth]{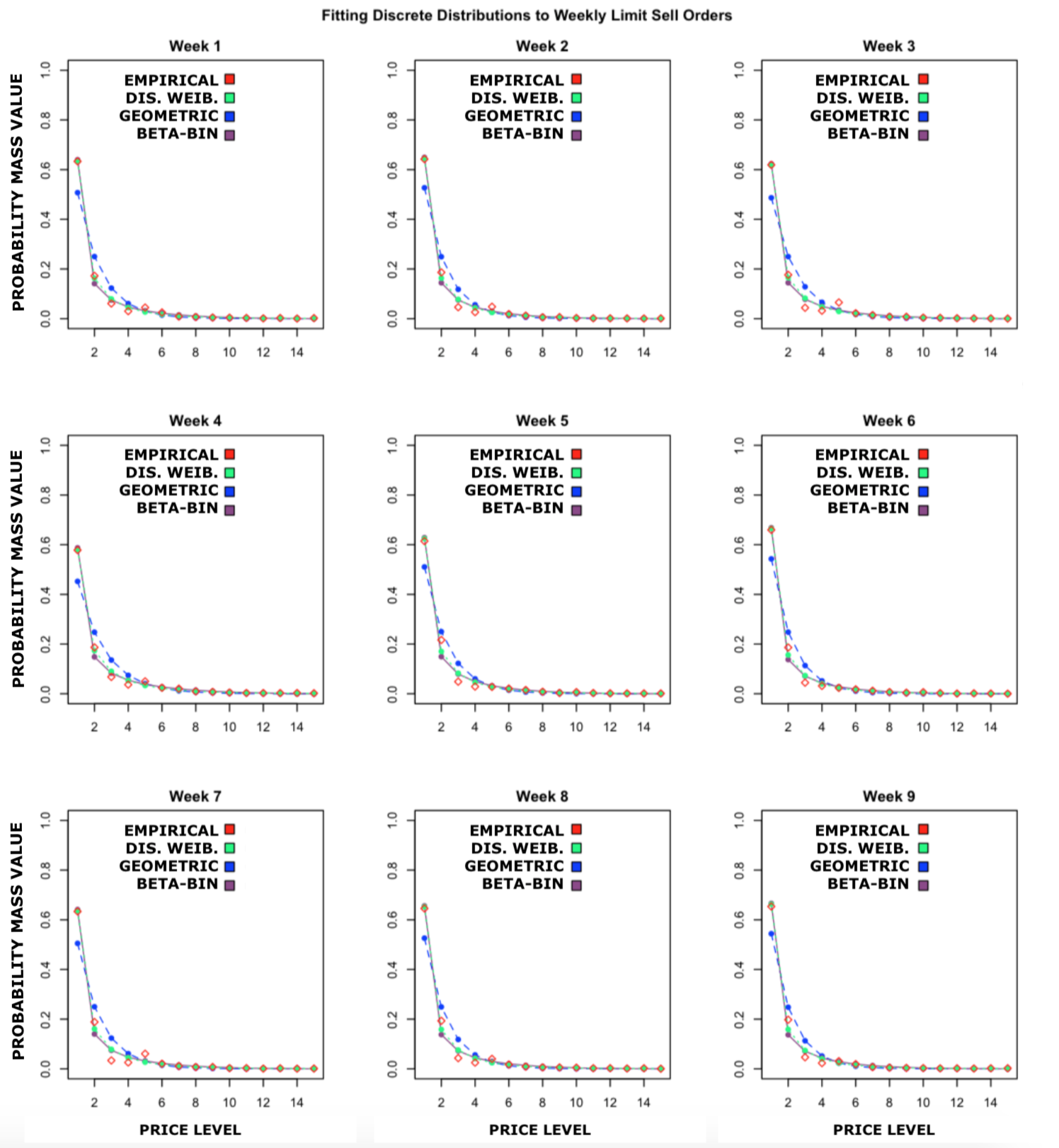}
    \caption{Discrete Fits on Weekly Limit Sell Arrival Rates}
    \label{fig:}
    
\end{figure}

We observe that Discrete Weibull distribution has the best fits for all of the weekly basis instances. Beta Binomial 
has close fit performance with respect to Discrete Weibull distribution. 
Geometric distribution is 3 to 4 times worse than both Discrete Weibull and Beta-Binomial distribution on average.

Geometric, Beta-Binomial and Discrete Weibull fits on monthly basis arrival rate of limit orders can be 
observed in Figure 5. We observe that Discrete Weibull distribution has the best fits for all of the monthly basis instances.
Geometric distribution is 3 to 4 times worse than others on average.

\begin{figure}[!ht]

    \centering
    \includegraphics[width=0.4\linewidth]{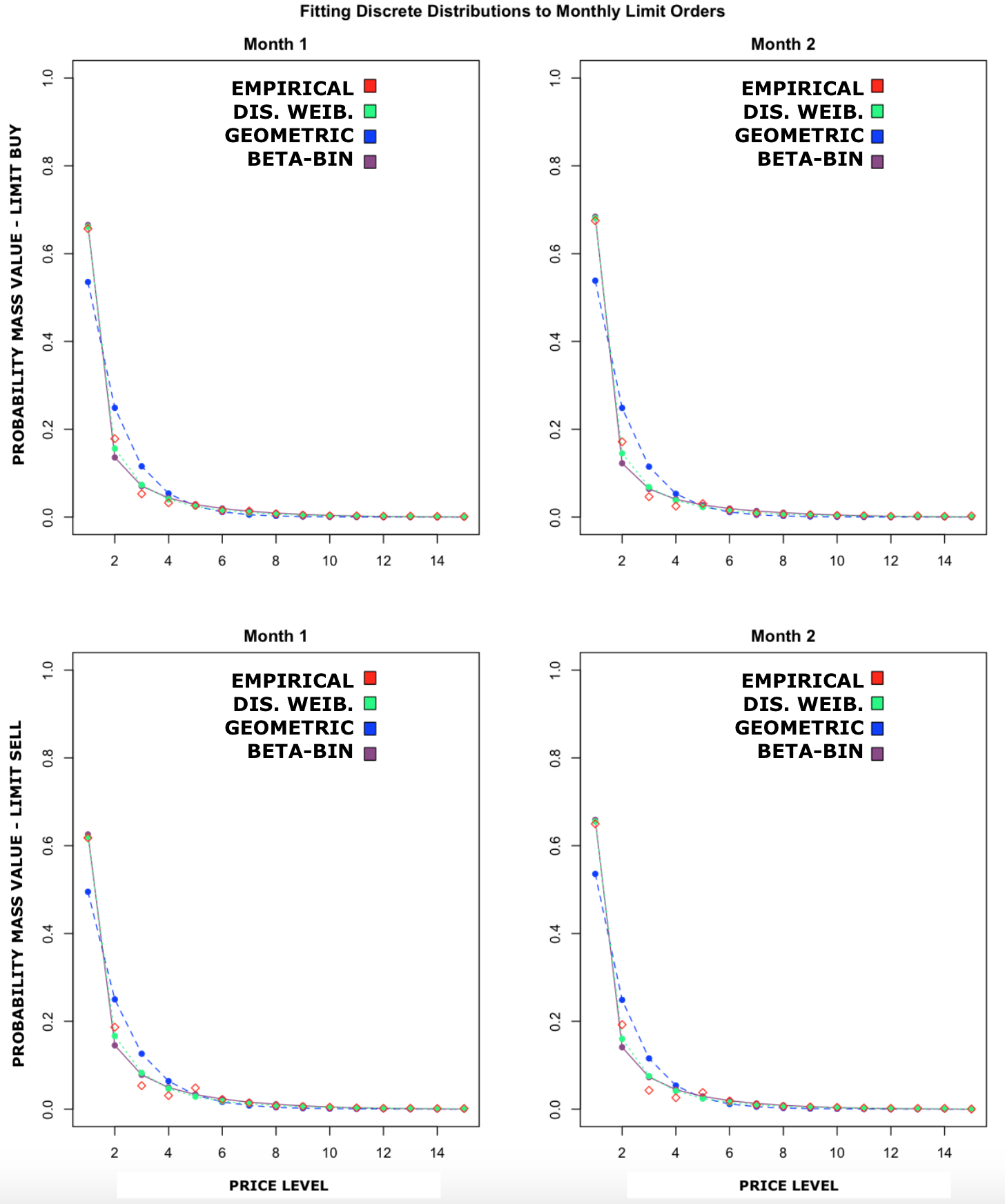}
    \caption{Discrete Fits on Monthly Limit Buy (above) and Sell (below) Arrival Rates}
    \label{fig:}
    
\end{figure}

Geometric, Beta-Binomial and Discrete Weibull fits on arrival rate of limit orders in hourly timesteps on different weeks  
can be observed in Figure 6 and Figure 7. Since there are 63 instances for this timestep, we only show the 
fits on arrival rates in the first 3 weeks and the first 3 hours. 

\begin{figure}[!ht]

    \centering
    \includegraphics[width=.93\linewidth]{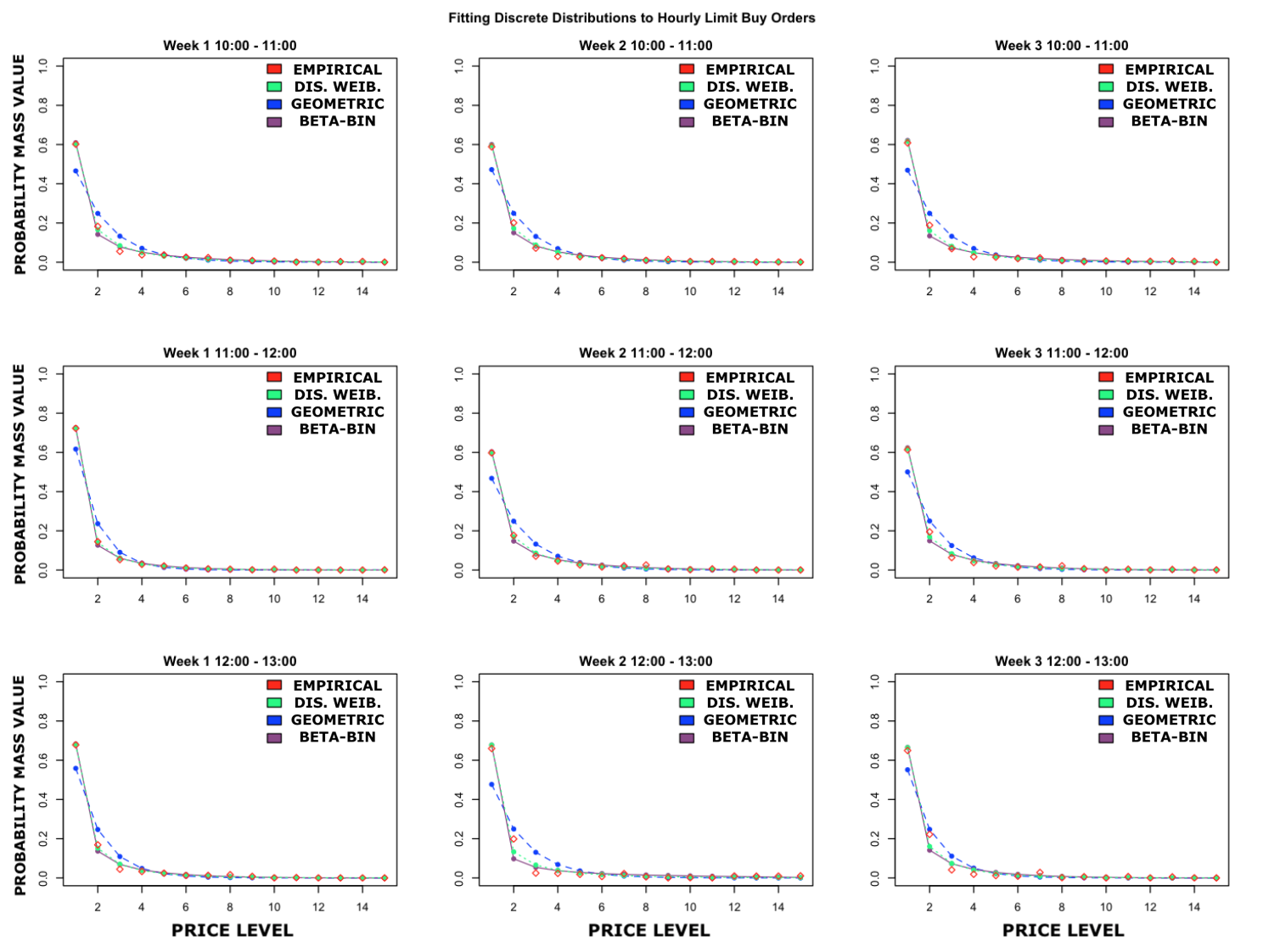}
    \caption{Discrete Fits on Hourly Limit Buy Arrival Rates}
    \label{fig:}
    
\end{figure}
\begin{figure}[!ht]

    \centering
    \includegraphics[width=.93\linewidth]{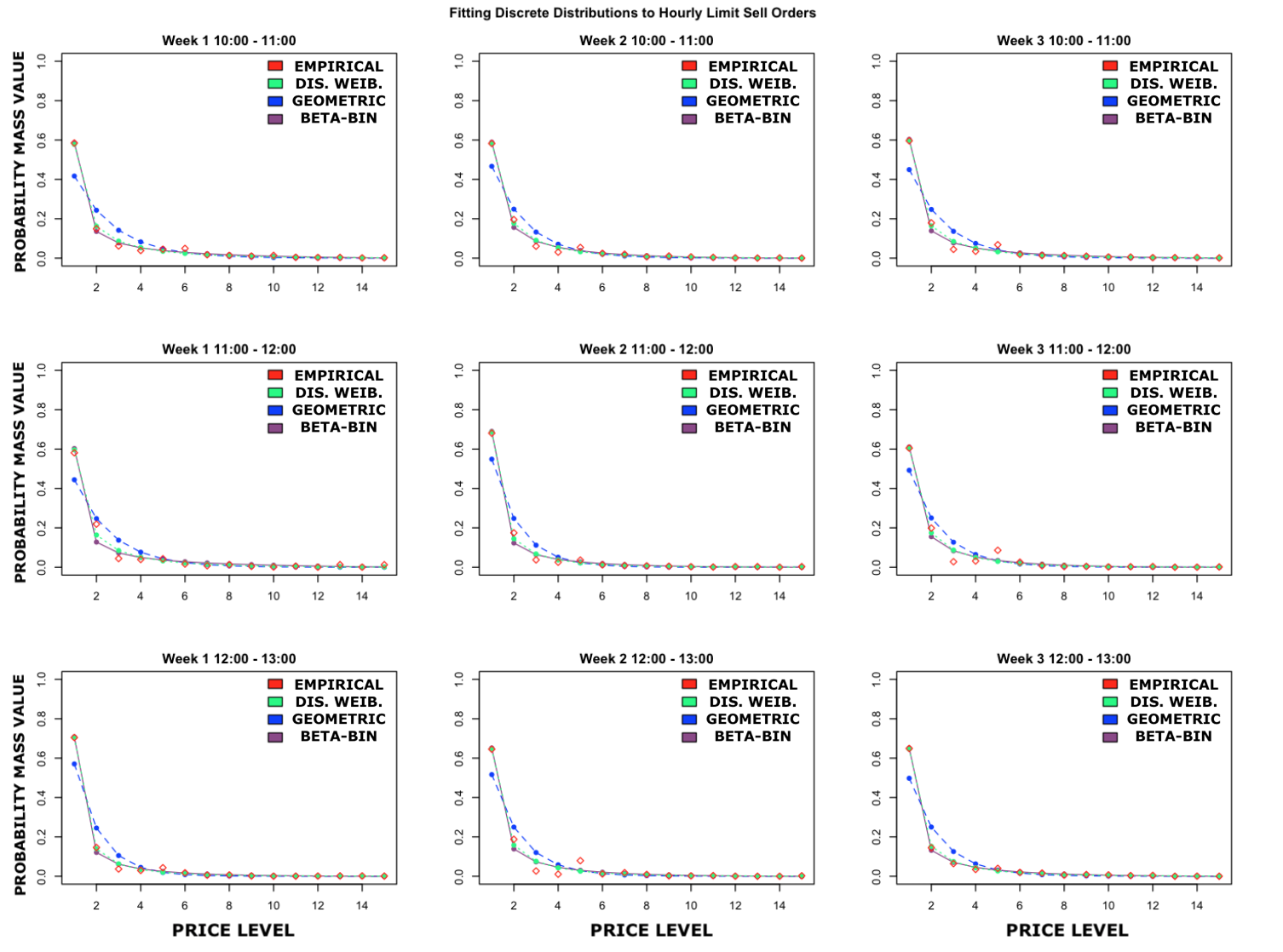}
    \caption{Discrete Fits on Hourly Limit Sell Arrival Rates}
    \label{fig:}
    
\end{figure}

\newpage

In order to compare the performance of discrete fits, we use the means of the performance scores of three distributions
in different timesteps. Welch's \textit{t}-test is used to decide which discrete distribution has the best fits on arrival 
rate of limit orders. Welch's \textit{t}-test is utilized to compare if there is a significant difference between two 
samples that have different variances. The means of performance scores are given in Table 1.

\begin{table}[!ht]
    \centering
    \begin{tabular}{l c c c}
    \hline
    \textbf{Timestep} & \textbf{Geometric} & \textbf{Discrete Weibull} & \textbf{Beta - Binomial} \\ [0.5ex] % inserts table %heading
    \hline
    Daily Limit Buy	& 4.042 +- 2.115 & 1.000 +- 0.000 & 1.262 +- 0.144 \\
    Daily Limit Sell & 3.144 +- 1.320 &	1.033 +- 0.126 & 1.214 +- 0.162 \\
    Weekly Limit Buy & 4.285 +- 0.779 &	1.000 +- 0.000 & 1.363 +- 0.106 \\
    Weekly Limit Sell &	3.183 +- 0.692 & 1.000 +- 0.000 & 1.237 +- 0.070 \\
    Monthly Limit &	3.789 +- 0.748 & 1.000 +- 0.000 & 1.312 +- 0.113 \\
    Hourly Limit Buy & 3.643 +- 1.760 & 1.012 +- 0.057 & 1.243 +- 0.160 \\
    Hourly Limit Sell &	2.895 +- 1.125 & 1.026 +- 0.083 & 1.153 +- 0.103 \\ [1ex]
    \hline
    \end{tabular}
    \caption{Mean $NPS$ of Discrete Fits on Different Timesteps}
\end{table}

It can be observed in Table 1 that the performance scores of Geometric fits have 3 to 4 times 
higher values than Discrete Weibull and Beta Binomial fits. As a result, we can say that 
Geometric fits have the worst performance among three distribution.
In order to find the best fits, we perform Welch's \textit{t}-test between Discrete Weibull and Beta Binomial fits.
The results of Welch's \textit{t}-tests as \textit{p}-values are given in Table 2.

\begin{table}[!ht]
    \centering
    \begin{tabular}{l c}
    \hline
    \textbf{Timestep} & \textbf{\textit{p}-value} \\ [0.5ex] % inserts table %heading
    \hline
    Daily Limit Buy	& 0.011 \\
    Daily Limit Sell & 0.033 \\
    Weekly Limit Buy & 0.003 \\
    Weekly Limit Sell &	0.020 \\
    Monthly Limit &	0.061 \\
    Hourly Limit Buy & 0.092\\
    Hourly Limit Sell &	0.039 \\ [1ex]
    \hline
    \end{tabular}
    \caption{Welch's \textit{t}-test results between Discrete Weibull and Beta-Binomial on Different Timesteps}
\end{table}

We choose 95\% confidence interval for \textit{t}-tests.
For 5 of 7 instances the \textit{p}-value is below 0.05, so we can reject the null hypothesis that indicates the means
of Discrete Weibull and Beta Binomial performance scores are not significantly different.
As a result, it can be denoted that Discrete Weibull has significantly better fits than Beta Binomial has in 
Daily Limit Buy, Daily Limit Sell, Weekly Limit Buy, Weekly Limit Sell and Hourly Limit Sell orders, 
since it has smaller means.
For Monthly Limit and Hourly Limit Buy orders, there is no significant difference between Discrete Weibull 
fits and Beta Binomial fits.

\subsection{Comparison of the Best Discrete Model and Theoretical Models on Arrival Rates of Limit Orders}

We perform least squares approach for power law parameter estimation, since it is suggested by \cite{bouchaud2002statistical}.
Error approach, timesteps and the comparison of performance scores are the same as in the discrete fits. 
We discretize Exponential fits by using the area under the probability density function. We divide the x-axis into 
15 equal parts and we find the densities for 15 ticks by calculating areas under the probability density function.
The limit buy and limit sell fits of Exponential, Power law and 
Discrete Weibull distributions from Day 29 to Day 40 can be seen in Figure 8 and Figure 9. 

\begin{figure}[!ht]

    \centering
    \includegraphics[width=\linewidth]{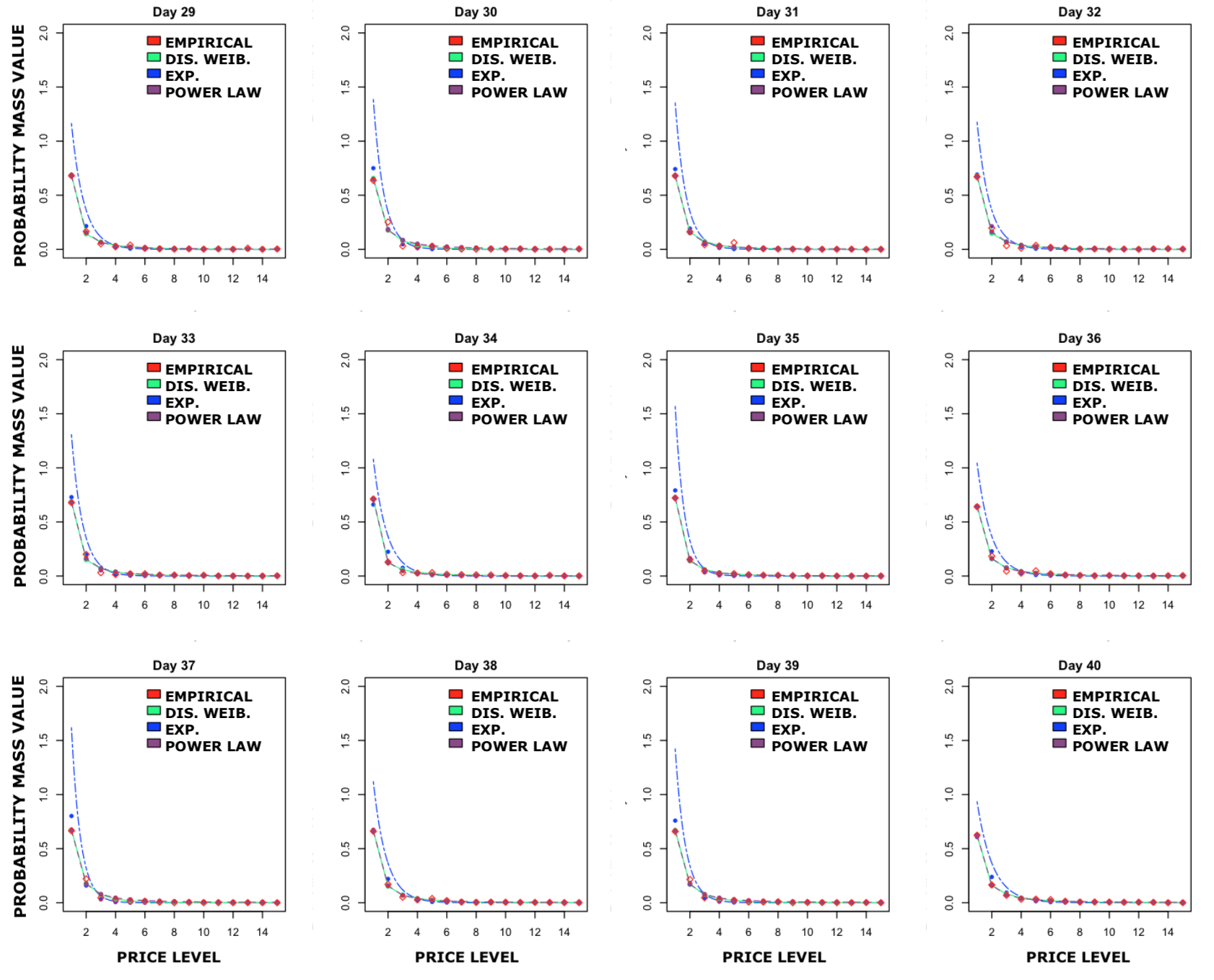}
    \caption{Exponential, Power law and Discrete Weibull fits on Daily Limit Buy Arrival Rates from Day 29 to Day 40}
    \label{fig:}
    
\end{figure}

\begin{figure}[!ht]

    \centering
    \includegraphics[width=.7\linewidth]{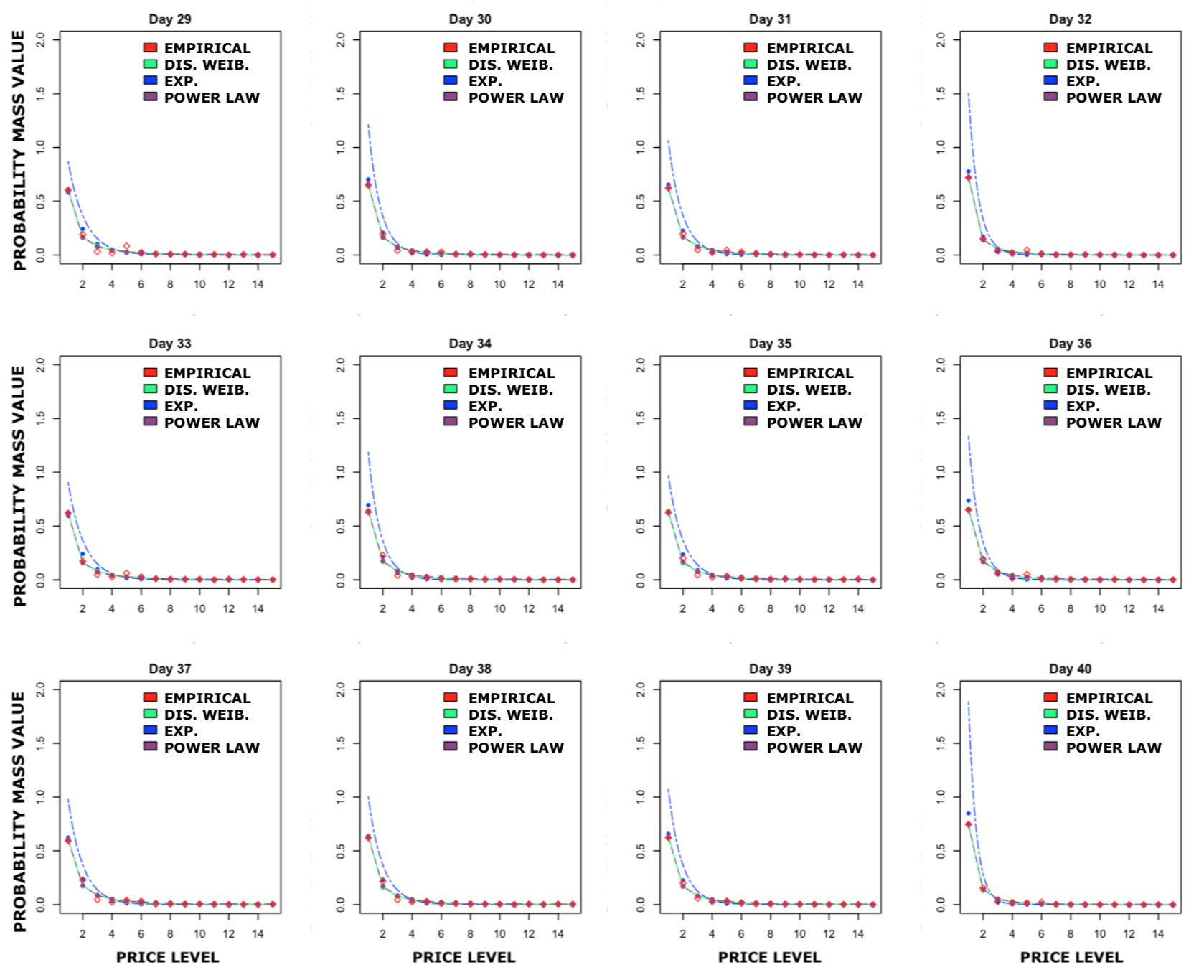}
    \caption{Exponential, Power law and Discrete Weibull fits on Daily Limit Sell Arrival Rates from Day 29 to Day 40}
    \label{fig:}
    
\end{figure}

We observe that Discrete Weibull distribution and Power law have the best fits for most of the instances. 
The performance of Discrete Weibull and Power law is very close to each other. On the other hand,
Exponential distribution is 2 to 3 times worse than both Discrete Weibull distribution and Power law. 
The limit buy and limit sell fits of Exponential, Power law and Discrete Weibull distributions from Week 1 to Week 9 can be seen in Figure 10
and Figure 11 below. 

\begin{figure}[!ht]

    \centering
    \includegraphics[width=.6\linewidth]{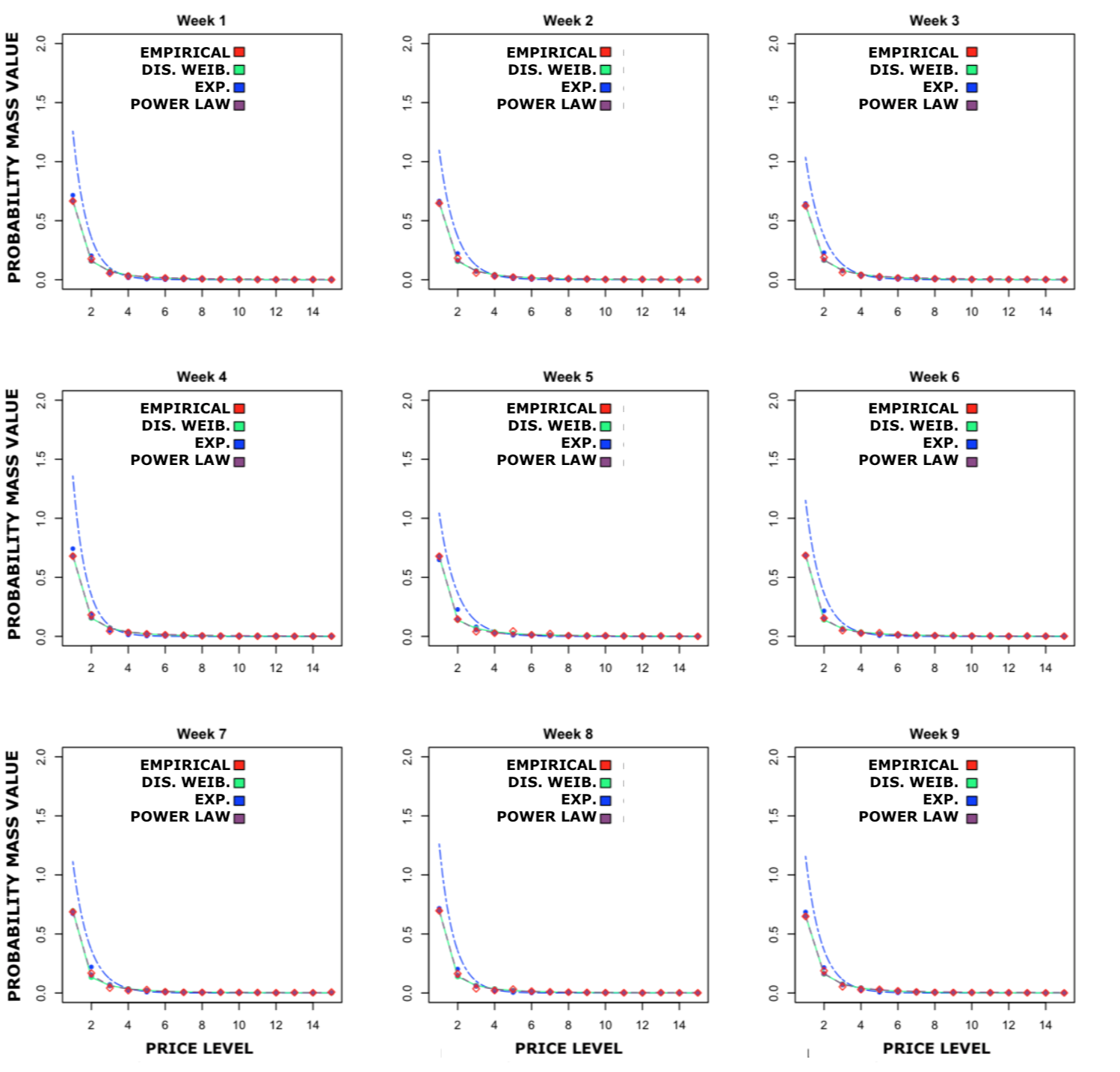}
    \caption{Exponential, Power law and Discrete Weibull fits on Weekly Limit Buy Arrival Rates}
    \label{fig:}
    
\end{figure}

\begin{figure}[!ht]

    \centering
    \includegraphics[width=.6\linewidth]{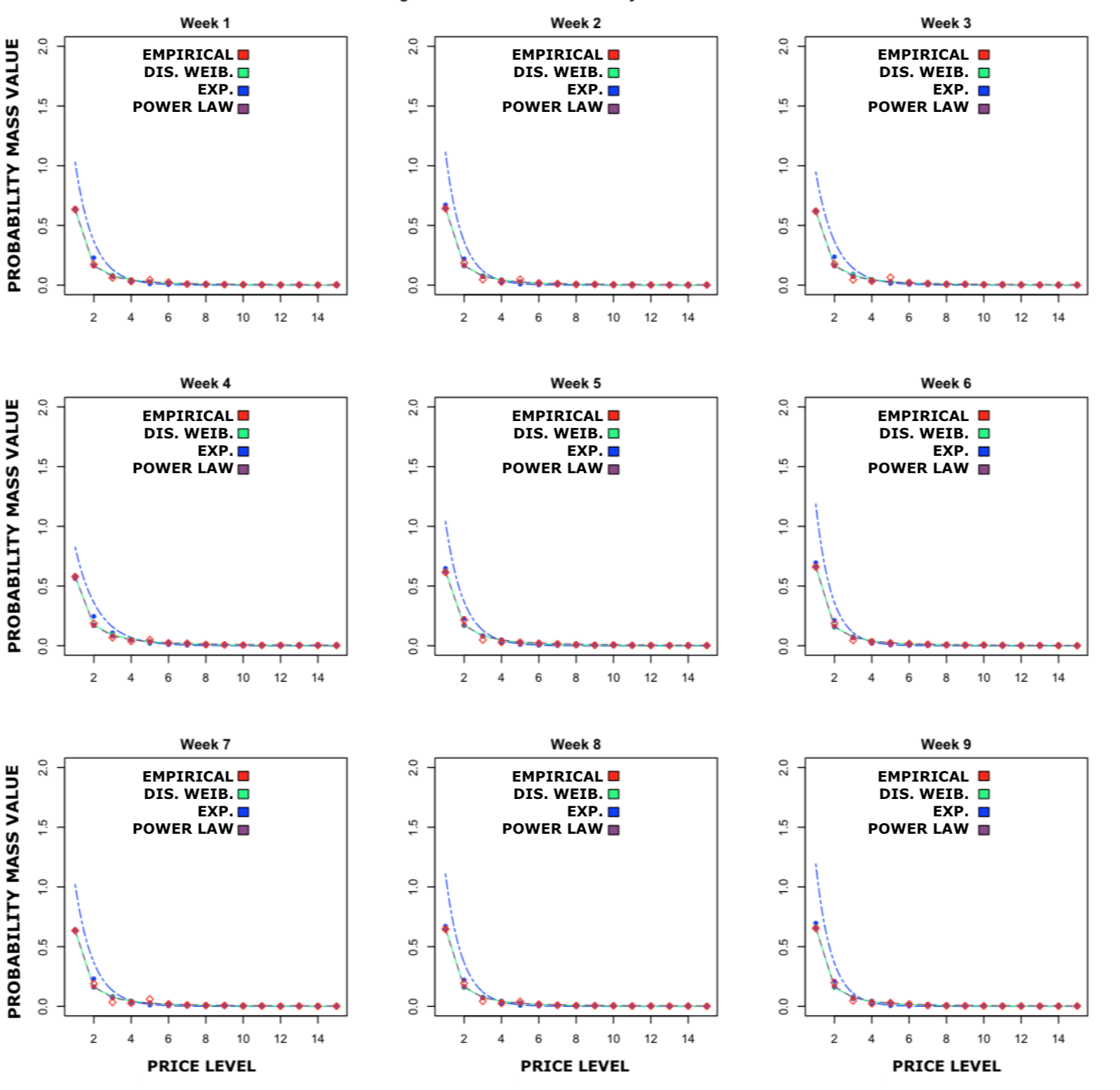}
    \caption{Exponential, Power law and Discrete Weibull fits on Weekly Limit Sell Arrival Rates}
    \label{fig:}
    
\end{figure}

We observe that Discrete Weibull distribution has the best fits for most of the weekly basis instances. Power law
has close fit performance with respect to Discrete Weibull distribution. 
Not suprisingly, Exponential distribution, being a more parsimonious distribution in the number of parameters, performed much worse than both Discrete Weibull and Power law on average.

Exponential, Power law and Discrete Weibull fits on monthly basis arrival rate of limit orders can be 
observed in Figure 12. 

\begin{figure}[!ht]
    \centering
    \begin{subfigure}[t]{0.5\textwidth}
        \centering
        \includegraphics[height=1.2in]{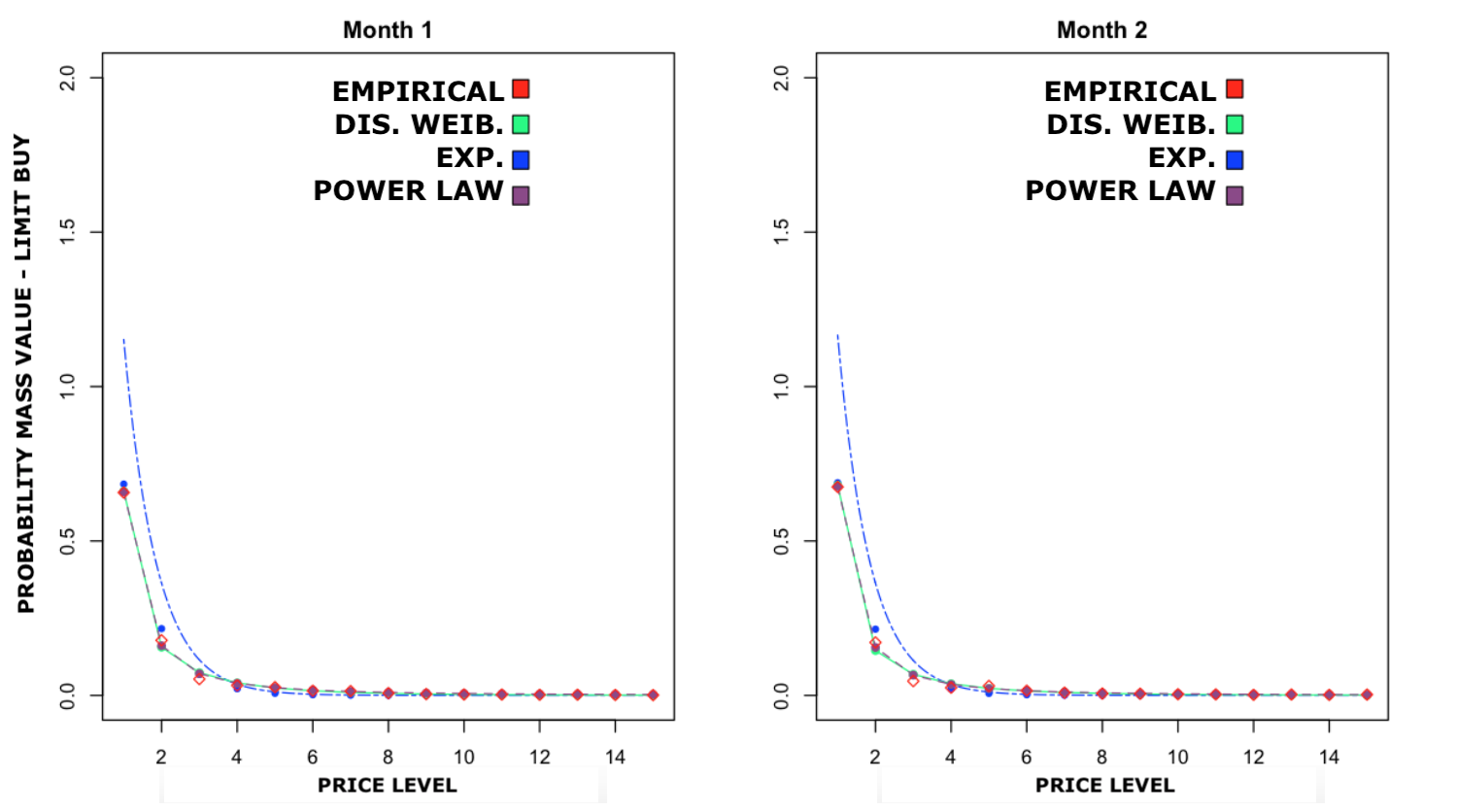}
        \caption{Limit Buy Orders}
    \end{subfigure}%
    ~ 
    \begin{subfigure}[t]{0.5\textwidth}
        \centering
        \includegraphics[height=1.2in]{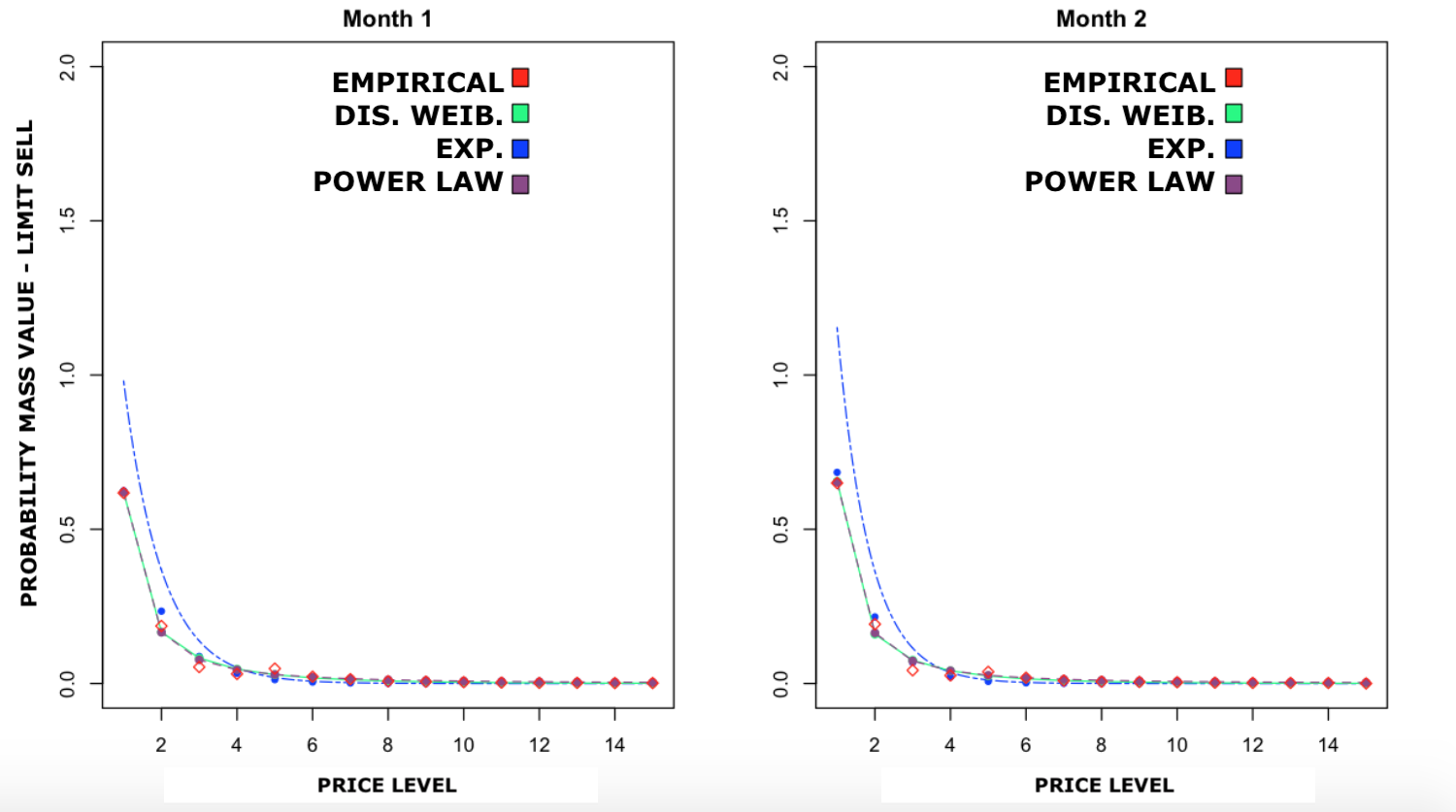}
        \caption{Limit Sell Orders}
    \end{subfigure}
    \caption{Exponential, Power law and Discrete Weibull fits on Monthly Limit Orders Arrival Rates}
\end{figure}

We observe that Discrete Weibull distribution has the best fits for 3 out of 4 monthly basis instances.
Exponential distribution is 2 times worse than both Discrete Weibull and Power law on average.

In order to compare the performance of Exponential, Power law and Discrete Weibull, we again use the means of the performance scores of three distributions
in different timesteps. Welch's \textit{t}-test is used to decide which discrete distribution has the best fits on arrival 
rate of limit orders. Average performance scores of Discrete Weibull, Exponential and Power law fits on 
different timesteps are given in Table 3.

\begin{table}[ht]
    \centering
    \begin{tabular}{l c c c}
    \hline
    \textbf{Timestep} & \textbf{Exponential} & \textbf{Discrete Weibull} & \textbf{Power law} \\ [0.5ex] % inserts table %heading
    \hline
    Daily Limit Buy	& 2.674 +- 1.346 & 1.098 +- 0.198 & 1.154 +- 0.222 \\
    Daily Limit Sell & 2.018 +- 1.071 & 1.059 +- 0.104 & 1.128 +- 0.167\\
    Weekly Limit Buy & 2.515 +- 0.651 & 1.165 +- 0.191 & 1.051 +- 0.066\\
    Weekly Limit Sell &	1.771 +- 0.471 & 1.004 +- 0.014 & 1.076 +- 0.060 \\
    Monthly Limit &	2.038 +- 0.394 & 1.063 +- 0.127 & 1.048 +- 0.062 \\
    Hourly Limit Buy & 2.624 +- 1.279 & 1.094 +- 0.126 & 1.163 +- 0.268 \\
    Hourly Limit Sell &	1.918 +- 0.753 & 1.056 +- 0.074 & 1.128 +- 0.168 \\ [1ex]
    \hline
    \end{tabular}
    \caption{Mean $NPS_d(Daily)$ of Proposed Distributions on Different Timesteps}
\end{table}

Exponential fits evidently have the worst performance among three distribution. On the other hand, as Power law and 
Discrete Weibull have very close mean performance scores, we  perform a $t$-test to find if there is a significant 
difference between those values.

\begin{table}[ht]
    \centering
    \begin{tabular}{l c}
    \hline
    \textbf{Timestep} & \textbf{\textit{p}-value} \\ [0.5ex] % inserts table %heading
    \hline
    Daily Limit Buy	& 0.363 \\
    Daily Limit Sell & 0.326 \\
    Weekly Limit Buy & 0.343 \\
    Weekly Limit Sell &	0.464 \\
    Monthly Limit &	0.973 \\
    Hourly Limit Buy & 0.743\\
    Hourly Limit Sell &	0.242 \\ [1ex]
    \hline
    \end{tabular}
    \caption{Welch's \textit{t}-test results between Discrete Weibull and Power law}
\end{table}

For all of the instances, the \textit{p}-values have a value that is higher than 0.05, so we can not reject the null 
hypothesis that indicates the mean performance scores of Discrete Weibull and Power Law is not significantly different.
So we can say that the performance of Power Law and Discrete Weibull fits are not significantly different 
for all instances. Consequently, Discrete Weibull models can compete with Power law models which are proposed 
by \cite{bouchaud2002statistical} and \cite{zovko2002power}. We can use Discrete Weibull distribution to 
model arrival rates of limit orders with respect to distance to the best prices accurately.

\subsection{Behavior of Order Cancellation Rates}

We analyze the number of cancel orders and the ratio of canceled orders in the vicinity of the best prices.
We consider both cancel buy orders and cancel sell orders on the weekly and monthly basis. In previous works, \cite{blanchet2013continuous} denoted that the cancellation activity is much higher in the close regions to the best bid and ask prices. The 
number of cancel orders arrived at the first 10 ticks on the weekly basis in our experiments are shown in Figure 13 and Figure 14.

\begin{figure}[!ht]

    \centering
    \includegraphics[width=.65\linewidth]{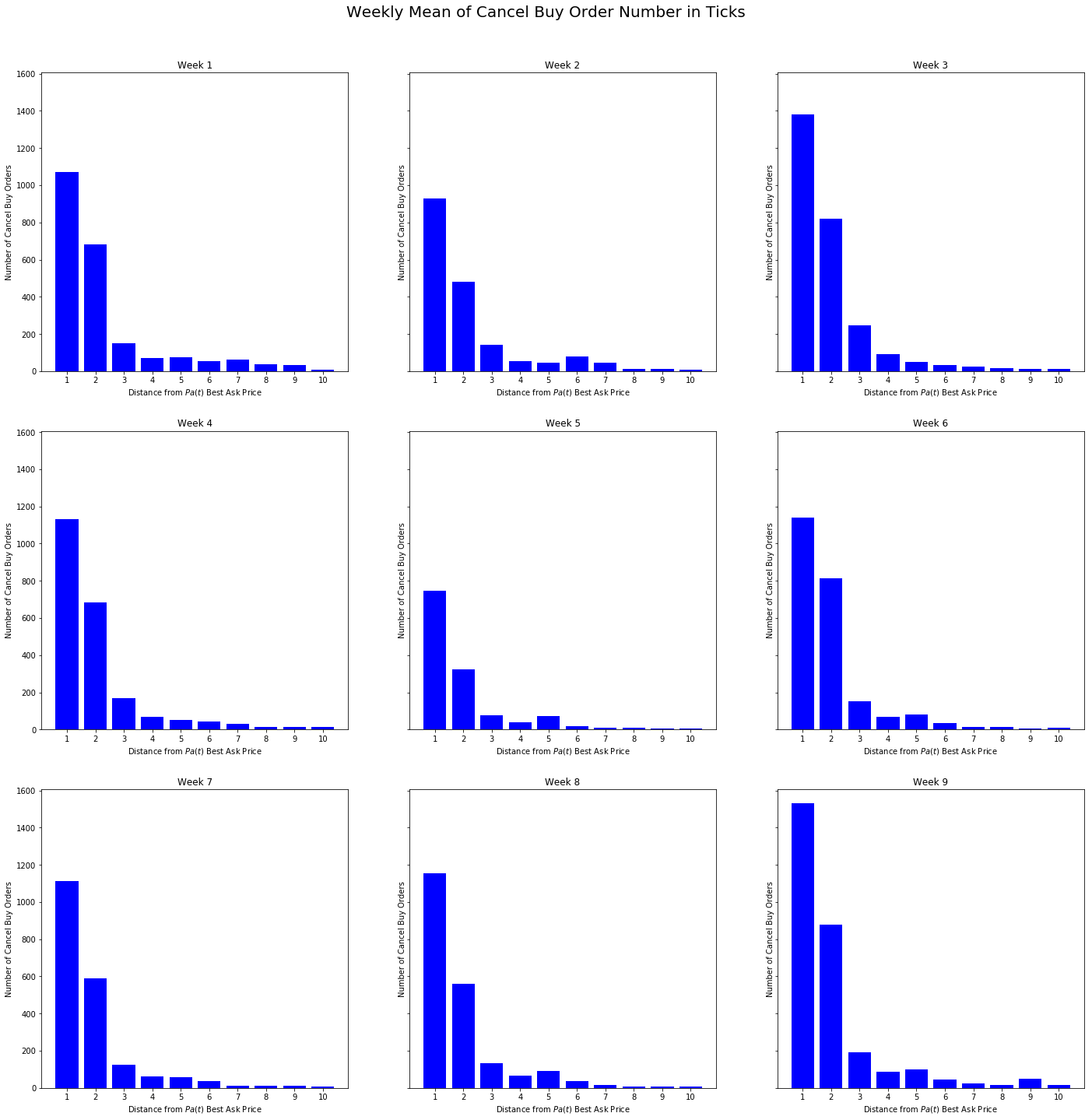}
    \caption{Number of Cancel Buy orders arrived in the vicinity of the best ask price on weekly basis}
    \label{fig:}
    
\end{figure}

\begin{figure}[!ht]

    \centering
    \includegraphics[width=.65\linewidth]{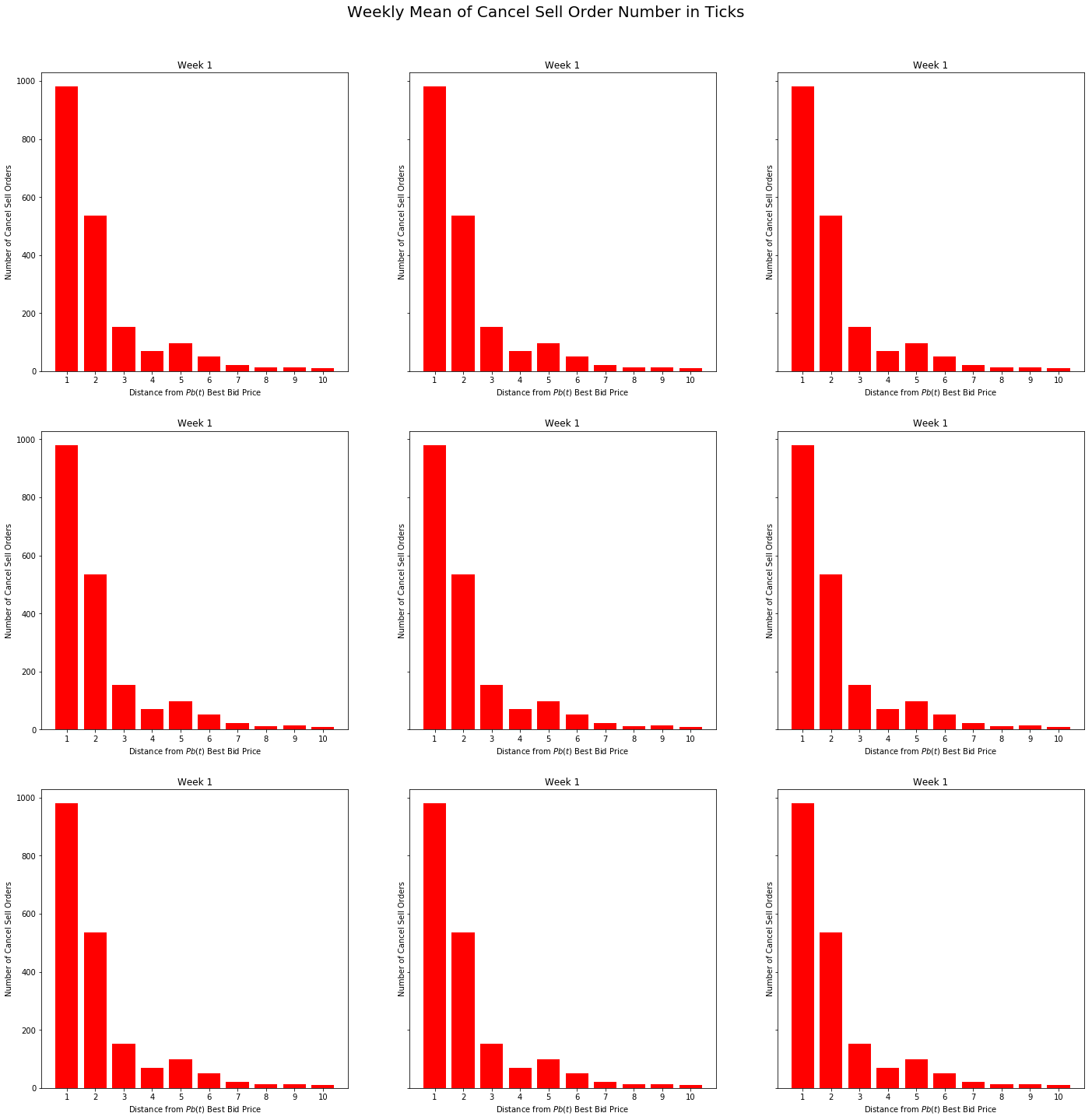}
    \caption{Number of Cancel Sell orders arrived in the vicinity of the best bid price on weekly basis}
    \label{fig:}
    
\end{figure}

When we consider the cancel activity as the number of cancel orders arrived, the results are consistent with 
the previous works. It can be observed that the number of cancel orders arrived in the close regions to the 
best prices are higher. We also consider the average ratio of canceled order quantity in the ticks. We use the metric
which we denote in Section 3.3 for finding the ratios. The 
ratios of canceled order quantities in the first 10 ticks on weekly basis in our experiments 
are shown in Figure 15 and Figure 16. The red and blue lines in figures expected values.

\begin{figure}[!ht]

    \centering
    \includegraphics[width=.65\linewidth]{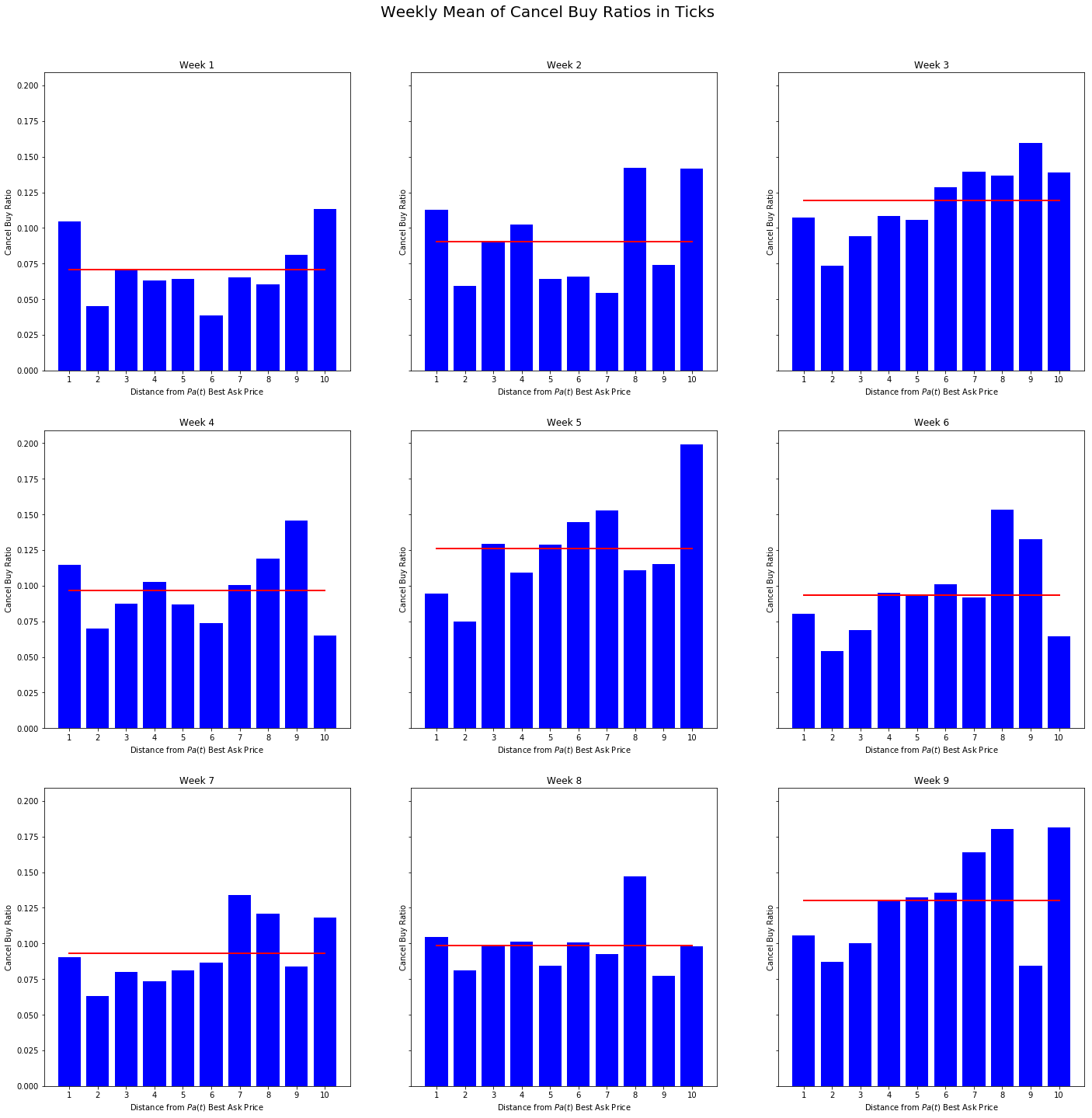}
    \caption{Ratios of canceled buy order quantities in the vicinity of the best ask price on weekly basis, Red line is the average ratio}
    \label{fig:}
    
\end{figure}

\newpage

\begin{figure}[!ht]

    \centering
    \includegraphics[width=.65\linewidth]{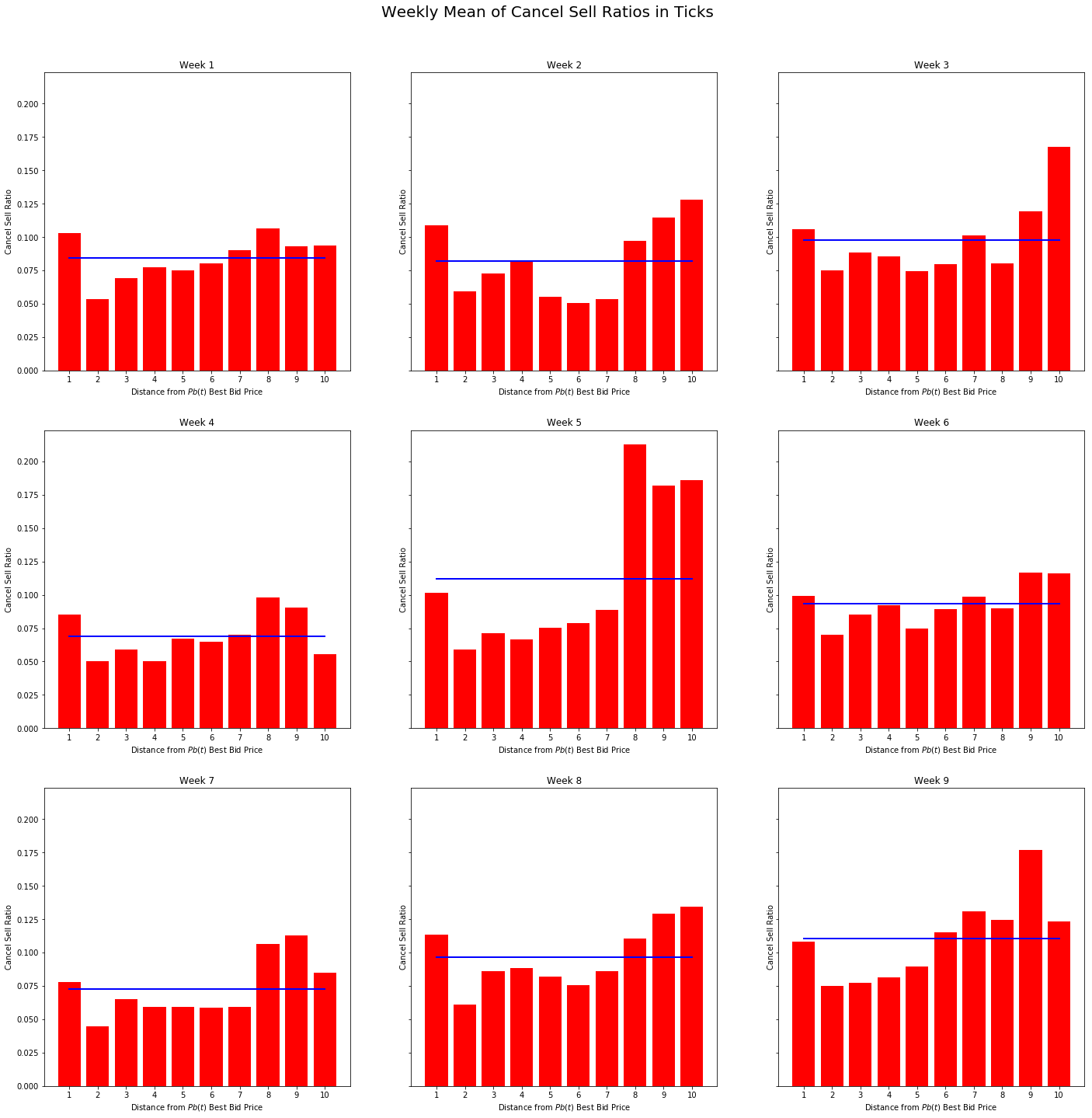}
    \caption{Ratios of canceled sell order quantities in the vicinity of the best bid price on weekly basis, Blue line is the average ratio}
    \label{fig:}
    
\end{figure}

The ratios of canceled order quantities in the first 10 ticks on monthly basis in our experiments 
are shown in Figure 17 and Figure 18.

\begin{figure}[!ht]

    \centering
    \includegraphics[width=.8\linewidth]{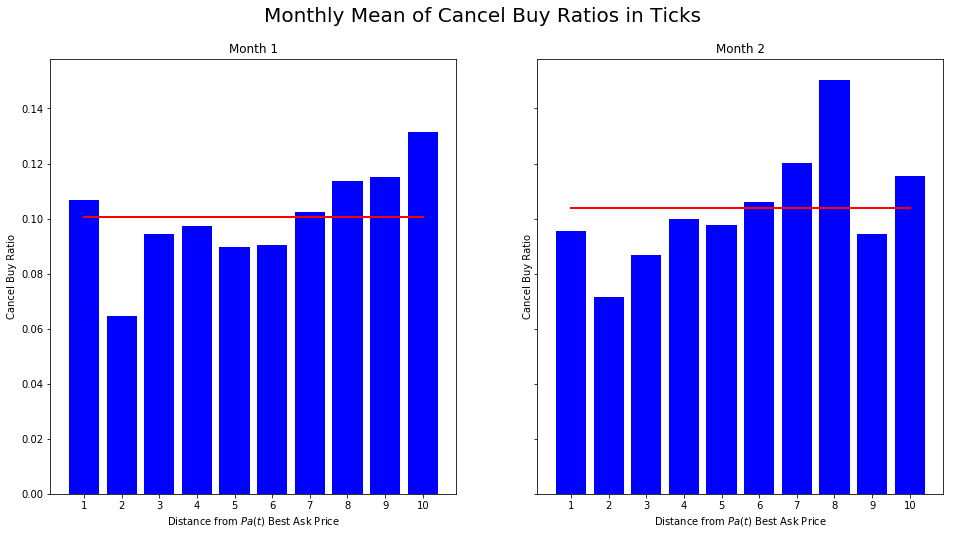}
    \caption{Ratios of canceled buy order quantities in the vicinity of the best ask price on monthly basis, Red line is the average ratio}
    \label{fig:}
    
\end{figure}

\begin{figure}[!ht]

    \centering
    \includegraphics[width=.8\linewidth]{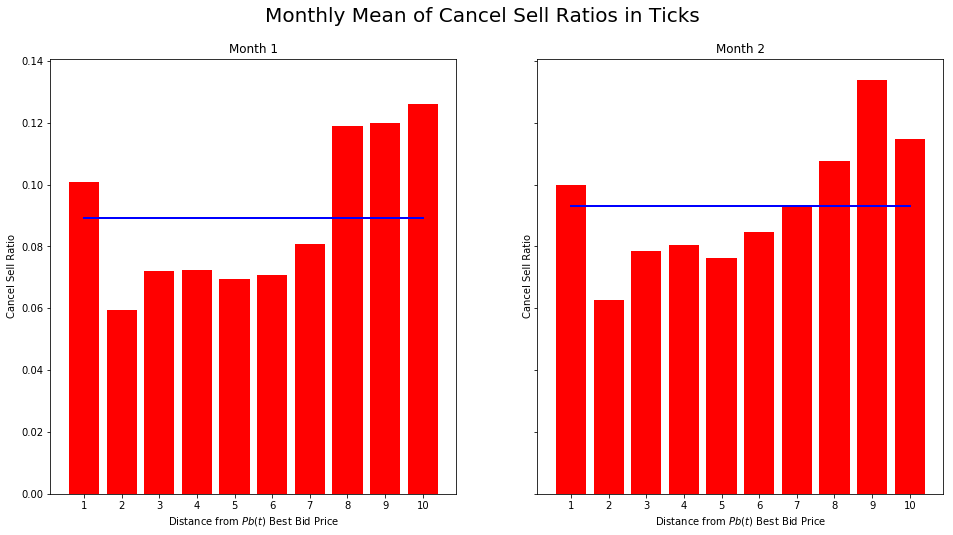}
    \caption{Ratios of canceled sell order quantities in the vicinity of the best bid price on monthly basis, Blue line is the average ratio}
    \label{fig:}
    
\end{figure}

The ratios were close most of the time. Thus we perform uniformity tests on the ratios. We use Chi-Square Test to test
if the ratios are distributed uniformly. Chi-Square test statistics formula is given in Equation 16. 
In the Equation 16, $c$ means the degree of freedom.
In our experiment, we have 10 different categories (because of 10 ticks). Since the degree of freedom is 
equal to the number of categories minus 1, the degree of freedom value is 9 for our experiments.

\begin{align}
    X_c^2 = \sum_{i = 1}^{10} \frac{(Observed_i - Expected_i)^2}{Expected_i}
\end{align}

Chi-Square tests use count data as observed data. Therefore we convert the cancellation ratios to 
integer values by multiplying them by 100. Then, we test the null hypothesis which claims that the ratios of canceled 
orders are consistent with Uniform distribution. 
Chi-Square test statistics and corresponding \textit{p}-values are given in Table 5 and Table 6.

\begin{table}[!ht]
    \centering
    \begin{tabular}{l c c}
    \hline
    \textbf{Timestep} & \textbf{Cancel Buy} & \textbf{Cancel Sell} \\ [0.5ex] % inserts table %heading
    \hline
    1st Week & 7.846 & 3.296 \\
    2nd Week & 12.197 & 8.371 \\
    3rd Week & 4.762 & 7.461 \\
    4th Week & 6.419 & 3.793  \\
    5th Week & 8.378 & 28.419  \\
    6th Week & 9.543 & 2.948  \\
    7th Week & 5.045 &  7.558  \\
    8th Week & 3.494 & 5.355  \\
    9th Week & 9.845 & 8.459 \\ 
    1st Month &	3.430 & 6.173  \\
    2nd Month &	4.740 & 4.539  \\ [1ex]
    \hline
    \end{tabular}
    \caption{Comparison between the ratios of canceled orders and the uniform distribution: Chi-Square Test Statistics on Different Timesteps}
\end{table}

\begin{table}[!ht]
    \centering
    \begin{tabular}{l c c}
    \hline
    \textbf{Timestep} & \textbf{Cancel Buy} & \textbf{Cancel Sell} \\ [0.5ex] % inserts table %heading
    \hline
    1st Week & 0.5497 & 0.9513 \\
    2nd Week & 0.2024 & 0.4972 \\
    3rd Week & 0.8545 & 0.5892 \\
    4th Week & 0.6973 & 0.9245  \\
    5th Week & 0.4965 & 0.0008 \\
    6th Week & 0.3887 & 0.9663  \\
    7th Week & 0.8303 & 0.5792  \\
    8th Week & 0.9414 & 0.8023  \\
    9th Week & 0.3631 & 0.4886 \\ 
    1st Month &	0.9447 & 0.7224  \\
    2nd Month &	0.8563 & 0.8725  \\ [1ex]
    \hline
    \end{tabular}
    \caption{Comparison between the ratios of canceled orders and the uniform distribution: Corresponding \textit{p}-values on Different Timesteps}
\end{table}

We use 95\% confidence interval for the tests. Chi-Square tests for uniformity present \textit{p}-values higher than 0.05 in all 
of the instances except for the ratios of cancel sell orders in the 5th week. Consequently, we can not reject the null hypothesis which indicates that the cancellation rates
are consistent with Uniform distribution for most of the instances.

\section{Conclusion and Future Work}
In this research, we used different statistical distributions to fit the limit order quantities arrived in the 
vicinity of the best bid and ask prices. The fits are made on Garanti Bank stock data for the period from August - 
September 2017. We analyzed the daily, weekly and monthly mean limit order quantities arrived at the first 15 
levels from the best prices. Also, we considered the weekly mean quantities of limit orders arrived in 7 
different time intervals in a day. We used total sum of $L_1$ norms between empirical density and fit results in 
the first 15 price levels of the bid and ask prices to evaluate the goodness of fits. 

We observed that Discrete Weibull and Beta Binomial distributions are almost 4 times better at fitting the order quantity data than Geometric 
distribution. We had 228 instances to fit and Discrete Weibull has the lowest $L_1$ norm for 210 of them. Beta 
Binomial fits the data with the lowest $L_1$ norm for 17 of the instances and Geometric distribution has the best fit 
for only one of the instances. Additionally, we used Exponential distribution to fit the same 228 instances. We 
found the probability mass values by calculating the areas of 15 bins under the Exponential probability density 
function using discretization. Then we obtained the sum of $L_1$ norms between empirical density and 15 probability 
mass values, and compared the goodness of Exponential distribution fits with Discrete Weibull distribution fits. 
We observed that Discrete Weibull fits the daily, weekly and monthly mean quantities two times better than 
Exponential distribution. Also, Discrete Weibull fits can compete with Power law fits which are proposed in early works. 

We analyzed the weekly and monthly mean ratio of cancel orders in the 
first 10 price levels. We conducted Chi-Square tests to test the uniformity. We observed 
that we can not deny the hypothesis which claims that the cancellation rates are consistent with Uniform distribution. As a result, we found out that the assumption 
made by \cite{cont2010stochastic} on cancellation rates which denotes that the cancellation 
rates are distributed exponentially can not be adapted to Turkish markets. 

Our dataset was quite small, since it only contains the data of 2 months. Also, we only 
consider Garanti Bank stock data. In future work, the same experiments can be conducted 
for larger datasets such as 6 months or 1 year. Moreover, stock data from other 
companies can be considered, and the relation between different stocks would be 
another interesting extension. Additionally, the relation between stock prices and 
arrival rates can be examined and predicted.

\bibliography{manuscript}

\section*{Appendix}

\newcommand{\hbAppendixPrefix}{A}
\renewcommand{\thefigure}{\hbAppendixPrefix.\arabic{figure}}
\setcounter{figure}{0}
\renewcommand{\thetable}{\hbAppendixPrefix\arabic{table}} 
\setcounter{table}{0}
\renewcommand{\theequation}{\hbAppendixPrefix\arabic{equation}} 
\setcounter{equation}{0}

\begin{figure}[H]
    \centering
    \begin{subfigure}[t]{.45\textwidth}
        \includegraphics[width=\linewidth]{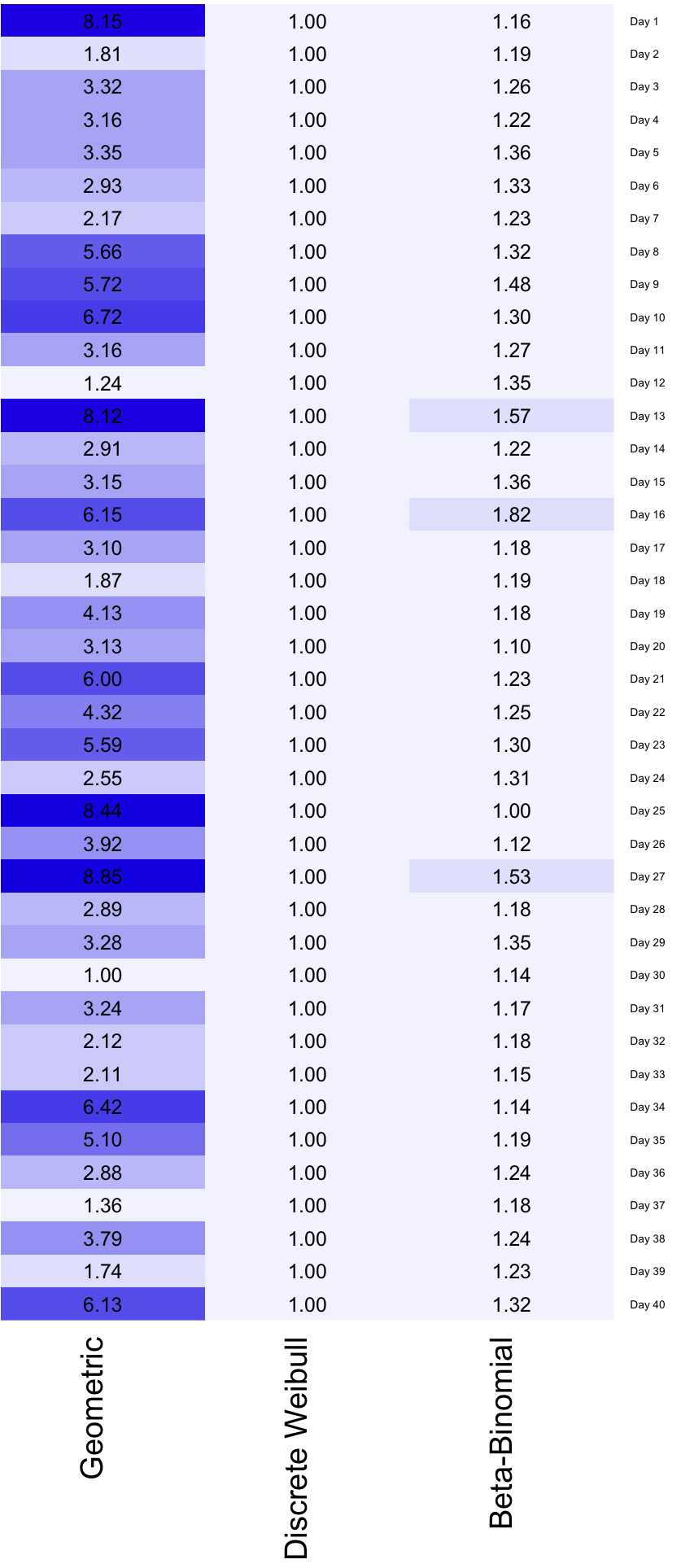}
    \end{subfigure}
    ~
    \begin{subfigure}[t]{.45\textwidth}
        \includegraphics[width=\linewidth]{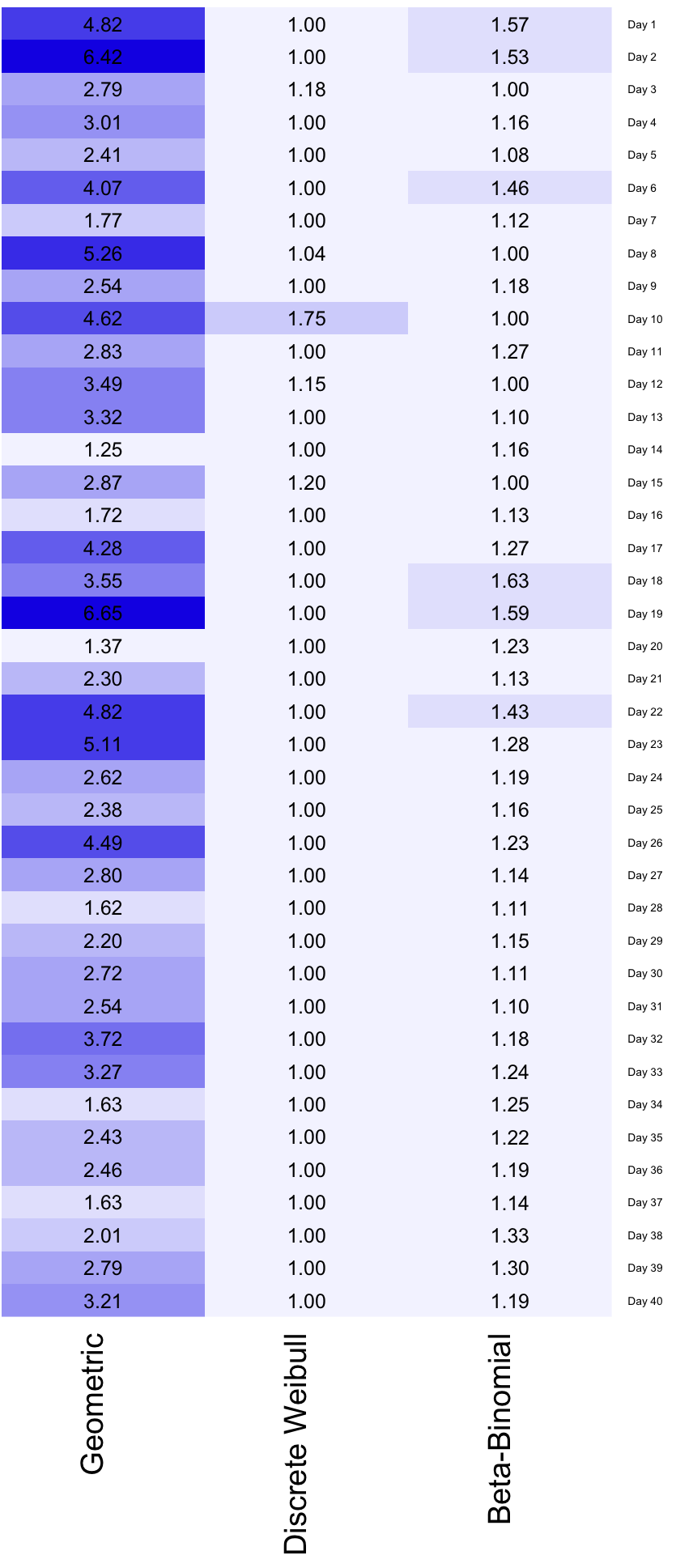}
    \end{subfigure}
    \caption[A.1]{$NPS_d(Daily)$ of three discrete distributions on Arrival Rates of Daily Limit Buy/Sell Orders}
\end{figure}

\begin{figure}[H]

    \centering
    \begin{subfigure}[t]{.45\textwidth}
        \includegraphics[width=\linewidth]{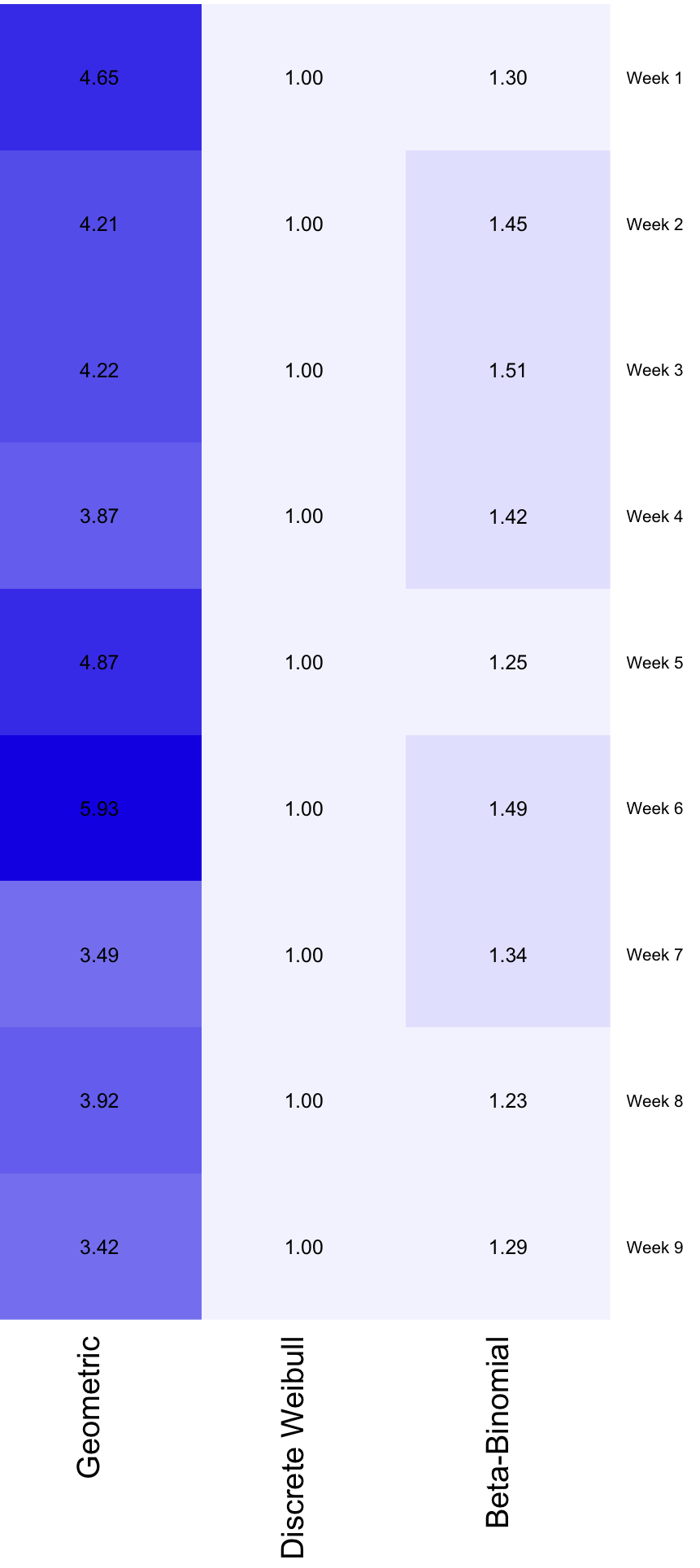}
    \end{subfigure}
    ~
    \begin{subfigure}[t]{.45\textwidth}
        \includegraphics[width=\linewidth]{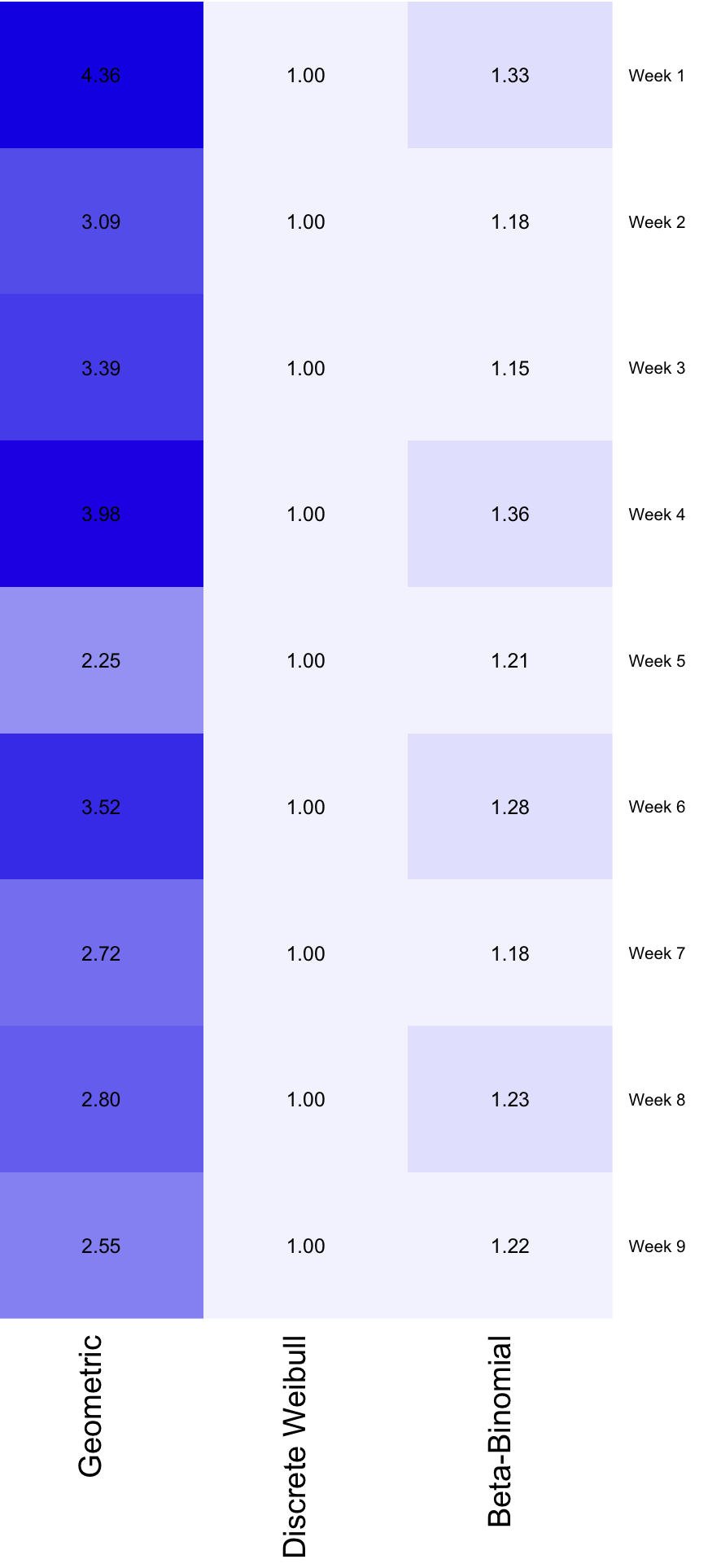}
    \end{subfigure}
    \caption{$NPS_d(Weekly)$ of three discrete distributions on Arrival Rates of Weekly Limit Buy/Sell Orders}%

\end{figure}

\begin{figure}[H]
    \centering
    \begin{subfigure}[t]{.45\textwidth}
        \includegraphics[width=\linewidth]{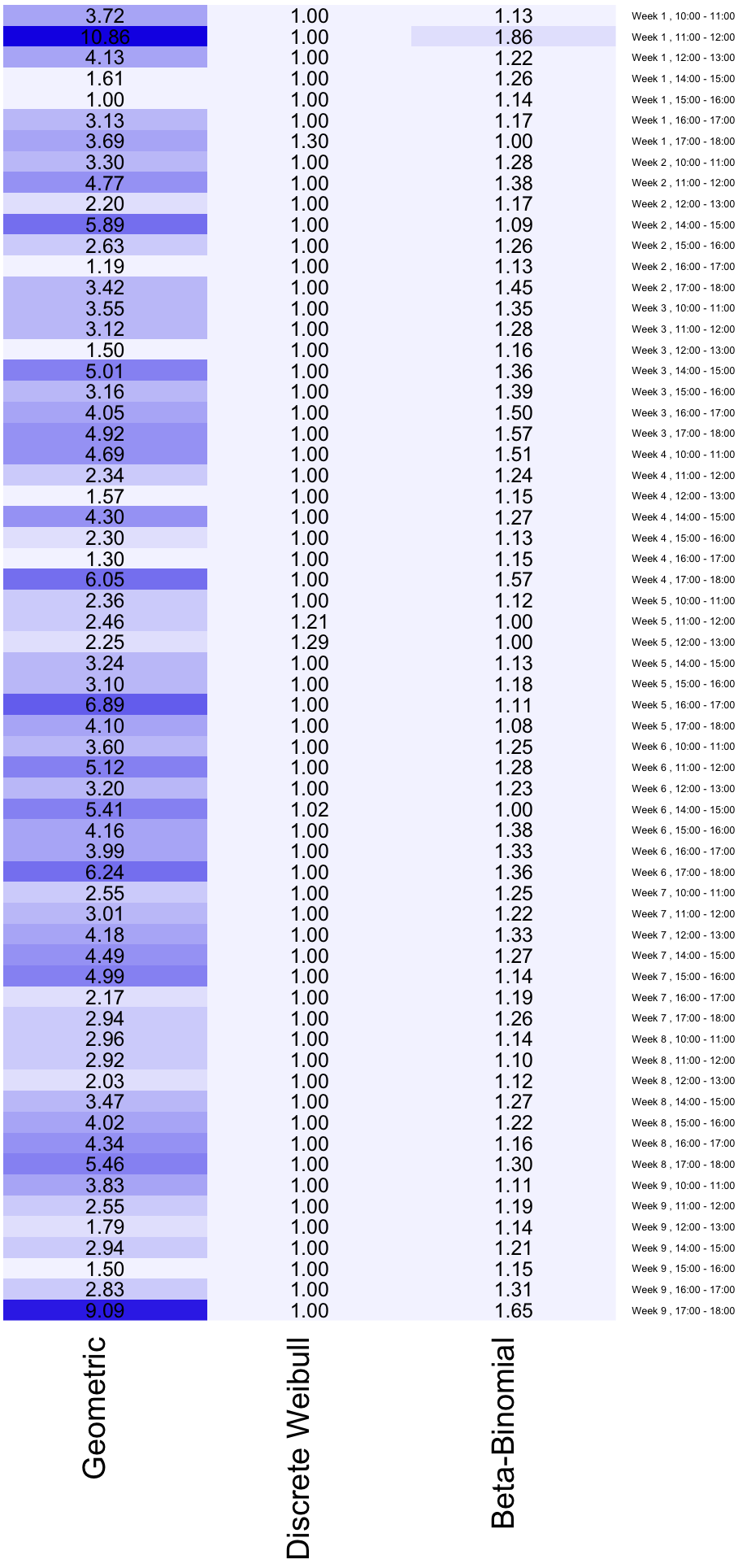}
    \end{subfigure}
    ~
    \begin{subfigure}[t]{.45\textwidth}
        \includegraphics[width=\linewidth]{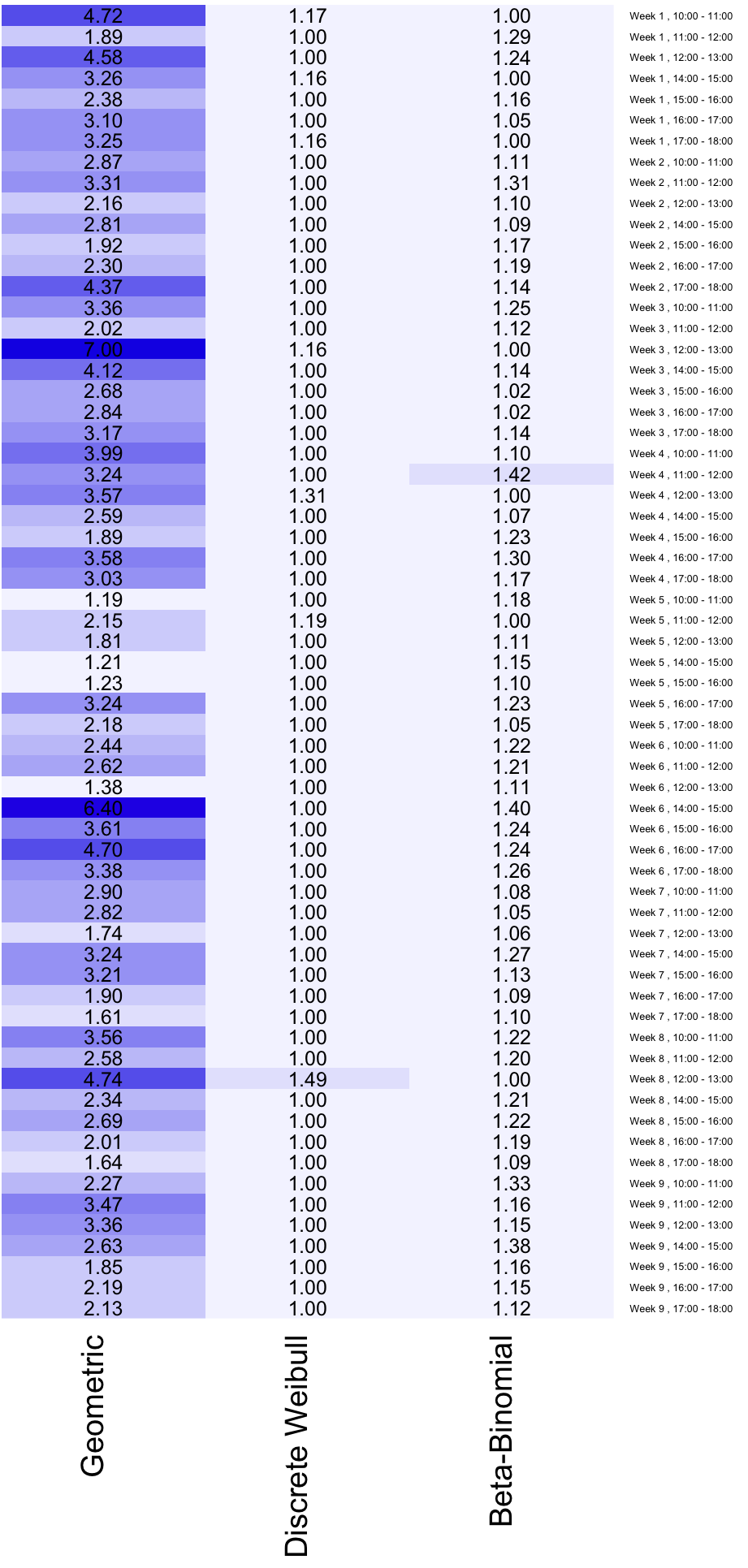}
    \end{subfigure}
    \caption{$NPS_d(Hourly)$ of three discrete distributions on Arrival Rates of Hourly Limit Buy/Sell Orders}%
\end{figure}

\begin{figure}[H]
    \centering
    \begin{subfigure}[t]{.45\textwidth}
        \includegraphics[width=\linewidth]{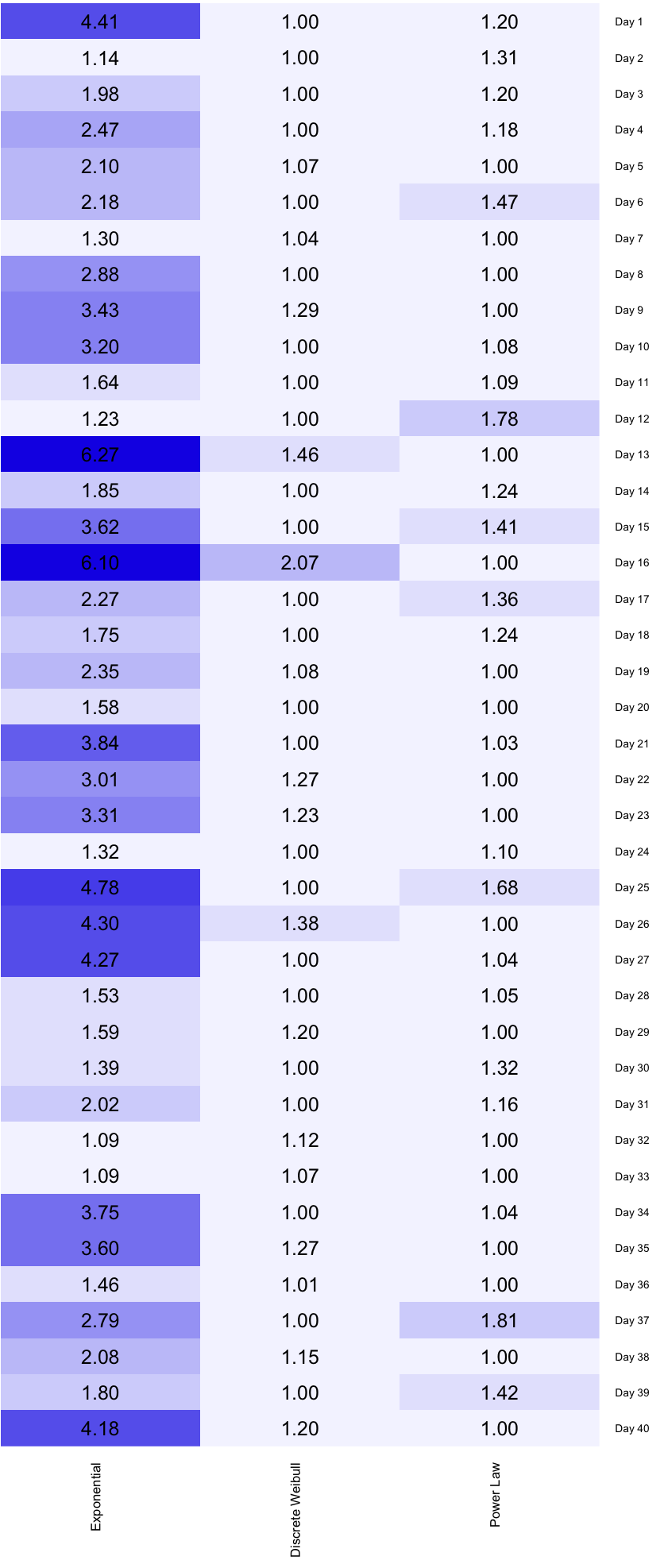}
    \end{subfigure}
    ~
    \begin{subfigure}[t]{.45\textwidth}
        \includegraphics[width=\linewidth]{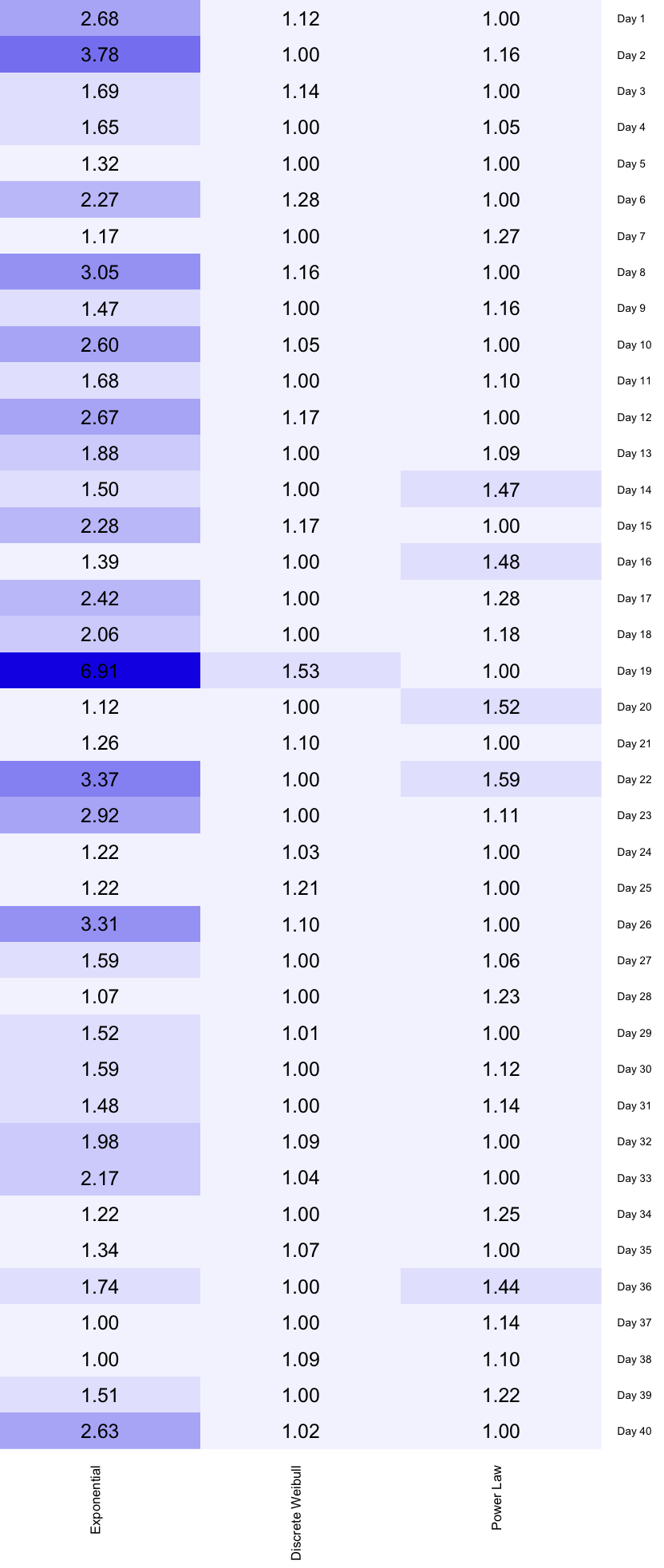}
    \end{subfigure}
    \caption[]{$NPS_d(Daily)$ of Discrete Weibull and theoretical distributions on Arrival Rates of Daily Limit Buy/Sell Orders}
\end{figure}

\begin{figure}[H]
    \centering
    \begin{subfigure}[t]{.45\textwidth}
        \includegraphics[width=\linewidth]{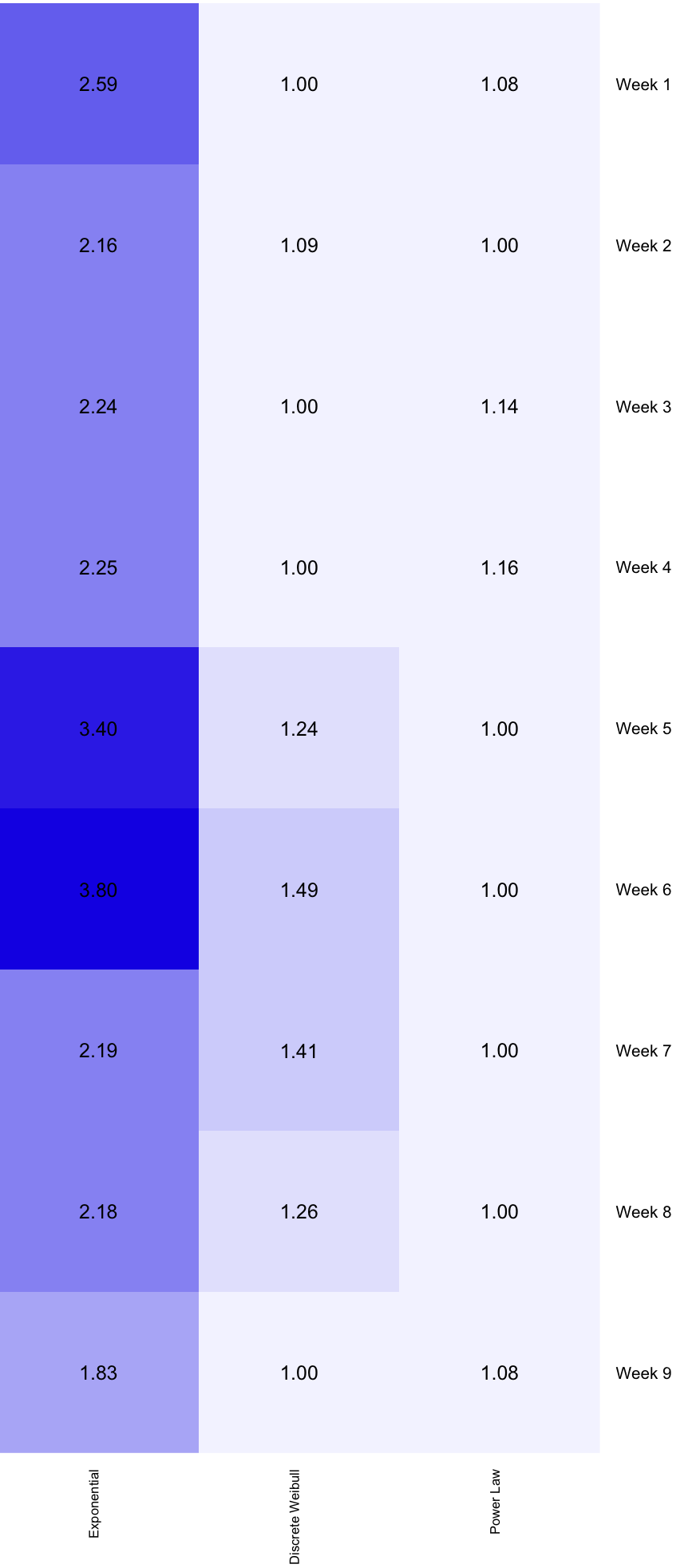}
    \end{subfigure}
    ~
    \begin{subfigure}[t]{.45\textwidth}
        \includegraphics[width=\linewidth]{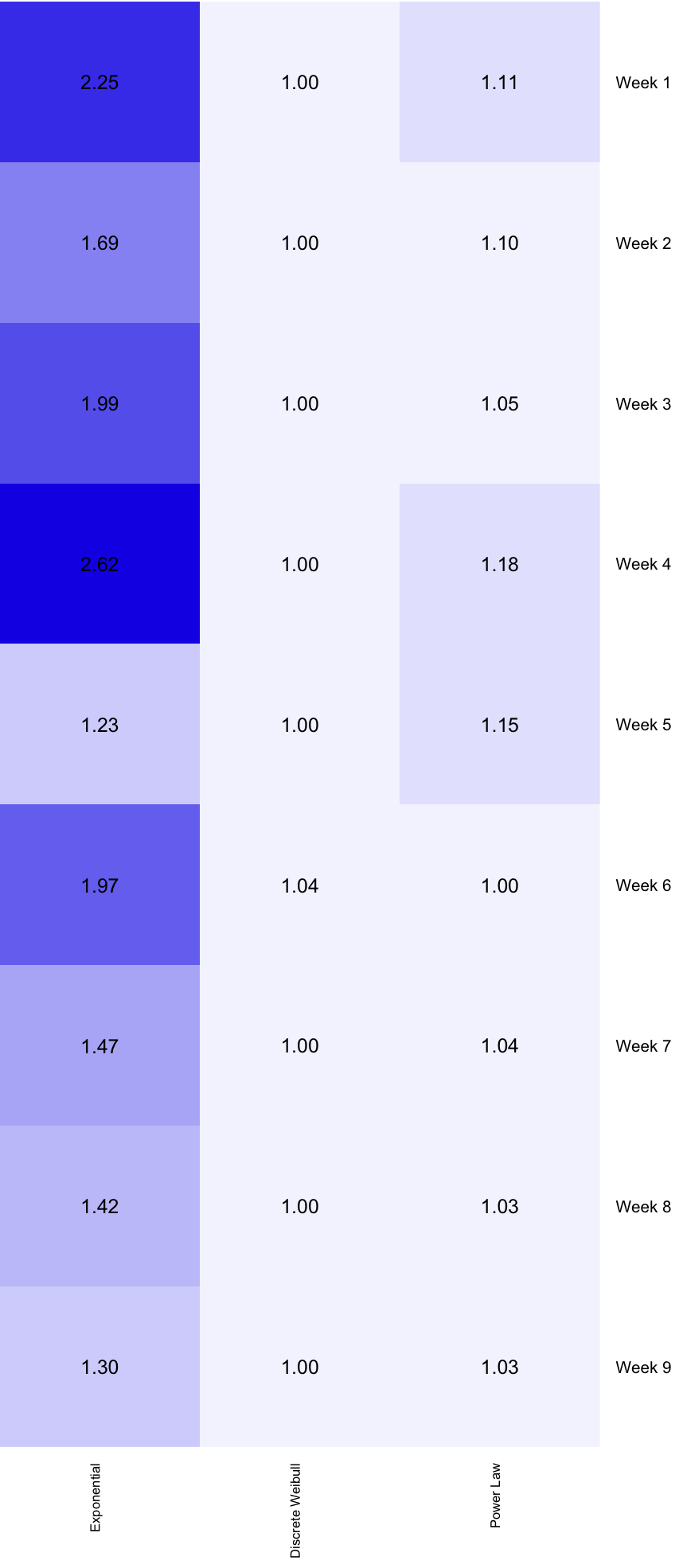}
    \end{subfigure}
    \caption[]{$NPS_d(Weekly)$ of Discrete Weibull and theoretical distributions on Arrival Rates of Weekly Limit Buy/Sell Orders}
\end{figure}

\begin{figure}[H]
    \centering
    \begin{subfigure}[t]{.45\textwidth}
        \includegraphics[width=\linewidth]{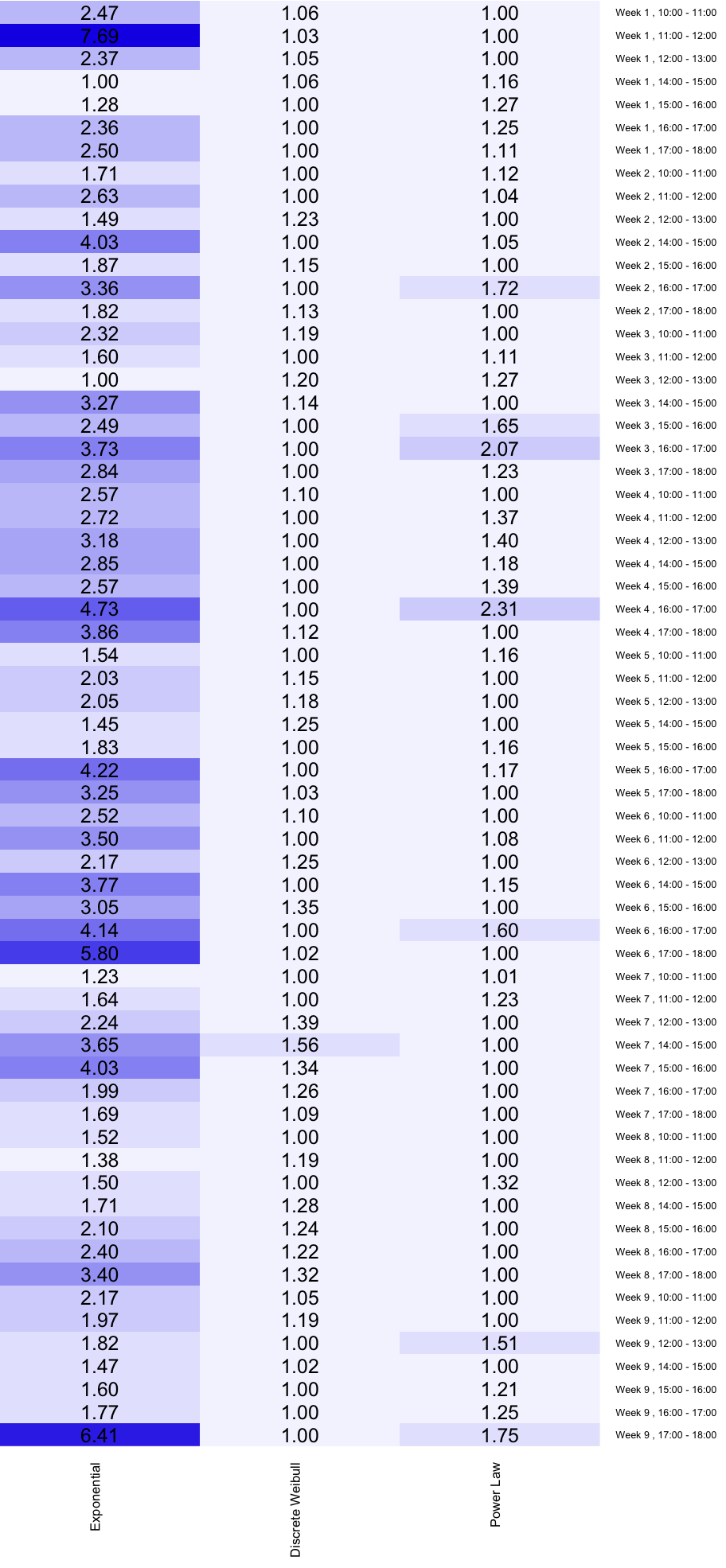}
    \end{subfigure}
    ~
    \begin{subfigure}[t]{.45\textwidth}
        \includegraphics[width=\linewidth]{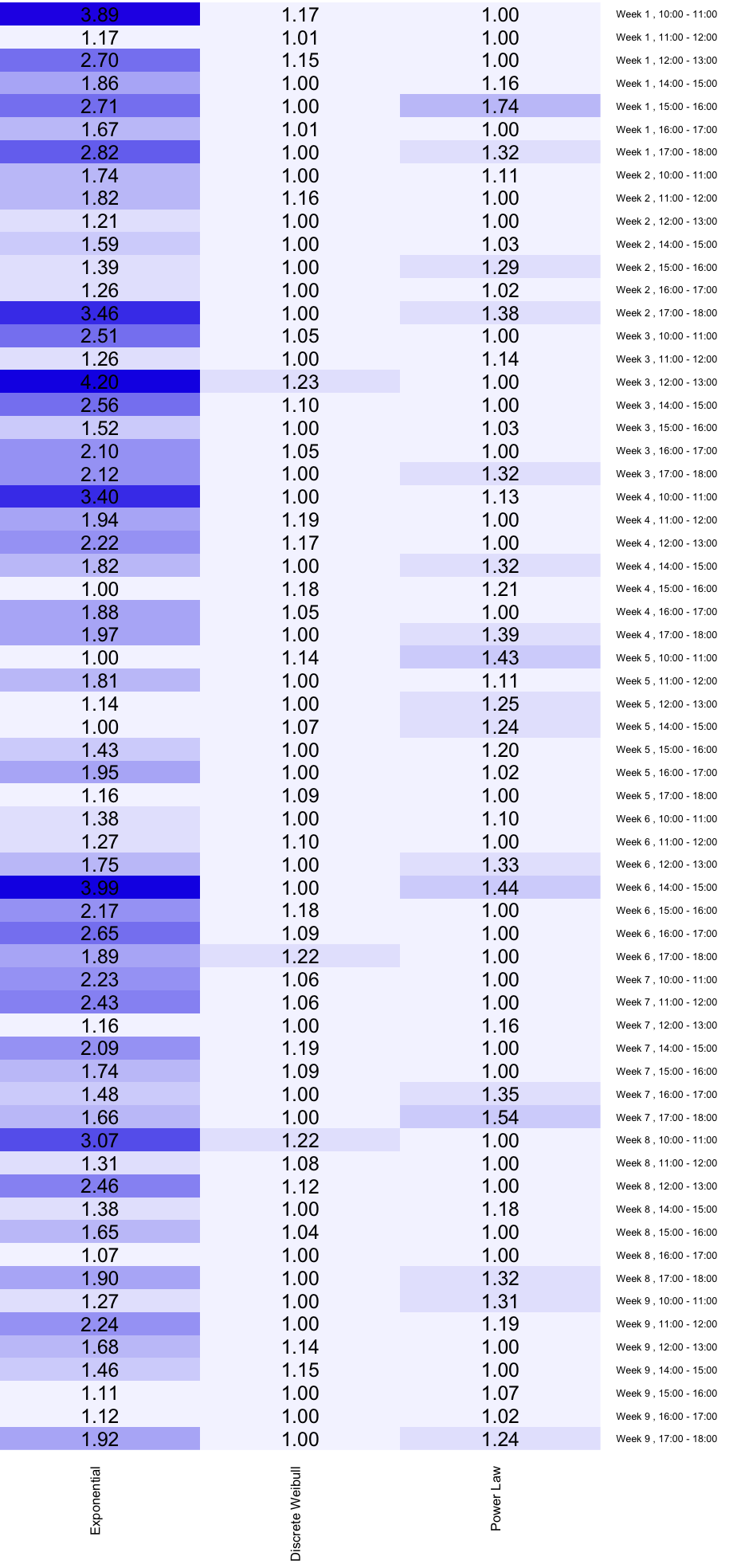}
    \end{subfigure}
    \caption{$NPS_d(Hourly)$ of Discrete Weibull and theoretical distributions on Arrival Rates of Hourly Limit Buy/Sell Orders}%
\end{figure}

\end{document}